\documentclass[12pt,preprint,aps,nofootinbib,shownopacs]{revtex4}
\usepackage{enumerate}
\usepackage{graphicx}
\newcommand*{\be}{\begin{equation}}
\newcommand*{\ee}{\end{equation}}
\newcommand*{\bea}{\begin{eqnarray}}
\newcommand*{\eea}{\end{eqnarray}}

\newcommand{\beq}{\begin{equation}}
\newcommand{\eeq}{\end{equation}}
\newcommand{\nn}{\nonumber}
\newcommand{\Frac}[2]{\frac{\displaystyle{#1}}{\displaystyle{#2}}}
\newcommand{\lsim}{\raise0.3ex\hbox{$\;<$\kern-0.75em\raise-1.1ex\hbox{$\sim\;$}}}
\newcommand{\gsim}{\raise0.3ex\hbox{$\;>$\kern-0.75em\raise-1.1ex\hbox{$\sim\;$}}}

\newcommand{\eq}[1]{Eq.~(\ref{#1})}

\newcommand{\BR}{{\cal B}}

\newcommand{\BXsga}{\bar{B} \to X_s \gamma}

\newcommand{\Btaun}{{B_u \to \tau \nu}}

\newcommand{\BRga}{{\cal B} (\BXsga)}

\newcommand{\mysigma}{\hspace{0.4mm} \sigma}

\newcommand{\CL}{{\cal L}}

\newcommand{\taunus} {\ensuremath{\bar{\tau}\nu}}

\newcommand{\ba}{\begin{array}}
\newcommand{\ea}{\end{array}}

\def\vevS{\langle S \rangle}
\def\vevFS{\langle F_S \rangle}

\def\Mess#1{M_{M_{#1}}}

\def\Mbar#1{{\overline{M}}_{#1}}
\def\Mhat#1{{\widehat{M}}_{#1}}

\def\mbar#1{{\overline{m}}_{#1}}

\def\alphbar#1{{\overline{\alpha}}_{#1}}

\def\mstau{m_{\widetilde \tau_R}}

\begin{document}

\preprint{IFIC-07-74, FTUV-07-1207}
\preprint{FERMILAB-PUB-07-643-T}
\title{Light charged Higgs at the beginning of the LHC era}
\author{Gabriela~Barenboim}
\email{gabriela.barenboim@uv.es}
\affiliation{Departament de F\'{\i}sica Te\`orica and IFIC, Universitat de
  Val\`encia-CSIC, E-46100, Burjassot, Spain.}

\author{Enrico~Lunghi}
\email{lunghi@fnal.gov}
\affiliation{Fermi National Accelerator Laboratory, P.O. Box 500, Batavia, IL
  60510-0500, USA.}
\author{Werner~Porod}
\email{porod@physik.uni-wuerzburg.de}
\affiliation{ Institut f\"ur Theoretische Physik und Astrophysik,
 Universit\"at W\"urzburg, D-97074 W\"urzburg, Germany.}

\author{Paride~Paradisi}
\email{paride.paradisi@uv.es}
\affiliation{Departament de F\'{\i}sica Te\`orica and IFIC, Universitat de
  Val\`encia-CSIC, E-46100, Burjassot, Spain.}

\author{Oscar~Vives}
\email{oscar.vives@uv.es}
\affiliation{Departament de F\'{\i}sica Te\`orica and IFIC, Universitat de Val\`encia-CSIC, E-46100, Burjassot, Spain.}
\begin{abstract}
  The terascale will be explored with the start of the LHC. One of the most
  fundamental questions which we expect to be answered is the root of
  electroweak symmetry breaking and whether the Higgs mechanism is realized in
  nature or not. In this context we pose the question if existing experimental
  data still allow for a light non-minimal Higgs sector.  We tackle this
  question first in the context of the two Higgs doublet model and then we
  concentrate in two supersymmetric models, the constrained MSSM and the MSSM
  with non-universal Higgs masses. In both supersymmetric scenarios, light
  pseudoscalar and light charged-Higgs bosons are still viable provided
  $\tan\beta$ is large. In this regime, we emphasize the importance of the
  constraints provided by the decay $B\to \tau \nu$ mediated by the
  charged-Higgs at tree-level. In addition we comment on generic predictions
  for hadronic colliders and indirect searches in such scenarios.
\end{abstract}

\maketitle

\section{Introduction}

The presence of a non-standard Higgs boson with a ``small'' mass, below 200 GeV,
would be a very interesting possibility in the first years of LHC operation.
In fact, the interest on this possibility has been recently increased with the 
small differences from Standard Model (SM) expectations found at CDF and 
D0 \cite{CDF,D0} and has motivated several analysis in the context of the
minimal supersymmetric standard model (MSSM) \cite{Ellis:2007ss,Feldman:2007fq}.
Even though the results are completely compatible with the absence of non-SM
Higgs bosons at the 2~$\sigma$ level these small discrepancies have
motivated the question whether it is possible to have a light non-SM Higgs
consistent with the present experimental constraints.
In this letter we intend to answer this question in models with 2 Higgs 
doublets and specially in the framework of the MSSM.
During the first years of LHC operation  and with the new measurements at
Tevatron, top quark physics will receive a big boost with a significantly
improved understanding of its physics and perhaps find a first clue of physics beyond the  SM.
 Perhaps the best possible situation to obtain sizeable 
beyond-the-SM effects in top-quark physics
corresponds to the existence of a charged Higgs boson of mass close to 
the top quark mass. In this work we will explore the possibility of having 
such a light Higgs sector in different models and how this affects 
phenomenology.

Clearly, the presence of a charged Higgs implies necessarily 
an extended Higgs sector. Therefore the simplest model we can explore and our
first option is a two Higgs doublet model (2HDM). In a generic type II 2HDM we see 
that the charged-Higgs is constrained to be heavier than 295 GeV by BR$(b\to s
\gamma)$, although a pseudoscalar mass in the range 150--200 GeV is still
allowed. As a second option we consider supersymmetric models, where we
find that a light charged-Higgs below 200 GeV is still possible both in the
Constrained MSSM (CMSSM) and in an MSSM with non-universal Higgs masses. 
However, in
these models the decay $B \to \tau \nu$ is a very strong constraint in the 
light $m_{H^+}$-large $\tan \beta$ region and, in particular, in the CMSSM sets
a strict lower limit of 180 GeV for the charged-Higgs mass. 

In the next section we explore in detail a generic type II two Higgs doublet model.
Section \ref{sec:MSSM} analyzes the CMSSM and a MSSM with non-universal 
Higgs masses and comments about models with mediation mechanisms other than gravity.
In section  \ref{sec:signals} we present the signatures of the light charged-Higgs scenario in collider and indirect search experiments. Finally in section  \ref{sec:conclusions} we present our conclusions.

\section{Two Higgs doublet models}
 The two Higgs doublet model is the simplest extension of the SM obtained with
the only addition of a second Higgs doublet. A 2HDM with generic Yukawa
couplings has severe Flavour Changing Neutral Currents (FCNC) problems  and,
for this reason, the Higgs couplings are restricted by an {\it ad hoc} discrete
symmetry to forbid FCNC at tree-level. The two main options are the type-I and
the type-II 2HDMs, depending on whether the up-type and down-type fermions are
coupled to the same or different Higgs doublets respectively. In our analysis,
we will assume a type-II 2HDM with a Higgs potential given by \cite{Gunion:1989we}
\begin{eqnarray}
  {V}_{\rm THDM}  &=&  m_1^2 \left| \Phi_1 \right|^2 
                          + m_2^2 \left| \Phi_2 \right|^2 - 
                              m_3^2 \left( \Phi_1^{\dagger} \Phi_2 
                              + \Phi_2^{\dagger} \Phi_1 \right) 
                                \nn  
                       + \frac{\lambda_1}{2} 
                               \left| \Phi_1 \right|^4 
                             + \frac{\lambda_2}{2} 
                               \left| \Phi_2 \right|^4 \nn \\
& &                          + \lambda_3 \left| \Phi_1 \right|^2 
                                \left| \Phi_2 \right|^2 
                      + \lambda_4 
                               \left| \Phi_1^{\dagger} \Phi_2 \right|^2
                             + \frac{\lambda_5}{2} 
                             \left\{ 
                               \left( \Phi_1^{\dagger} \Phi_2 \right)^2
                            +  \left( \Phi_2^{\dagger} \Phi_1 \right)^2
                             \right\},  \label{pot}
\end{eqnarray}
where $\Phi_i$ are the Higgs iso-doublets with hypercharge $\frac{1}{2}$.
Being a type-II 2HDM, this potential satisfies a (softly-broken) discrete 
symmetry under the transformation $\Phi_1\to\Phi_1$ and $\Phi_2\to-\Phi_2$.
A nonzero value of $m_3^2$ indicates that the discrete symmetry
is broken softly  and would correspond to a $B\mu$ coupling in supersymmetric 
models. The eight free parameters ($m_1^2$-$m_3^2$ and $\lambda_1$-$\lambda_5$) 
can be rewritten in terms of eight ``physical'' parameters, i.e. four Higgs mass 
parameters $m_h, m_H, m_A, m_{H^\pm}$, two mixing angles $\alpha$, $\beta$,
the vacuum expectation value $v$, and the soft-breaking scale of the discrete 
symmetry $M$. The two physical CP-even fields $h$ and $H$ are such that 
$m_h^{0} \le m_H^{0}$. In particular, the masses of the CP-odd, $A$ (CP-odd) and 
the charged Higgs, $H^\pm$, are related by the following expression
\begin{eqnarray}
m_{H^\pm}^2 -m_{A}^2 &=& \frac{1}{2} (\lambda_5-\lambda_4) v^2\, ,
\end{eqnarray}
 with $v = \sqrt{v_1^2 + v_2^2}$, where $v_1$ and $v_2$ are the vacuum
expectation values of $\Phi_1$ and $\Phi_2$, respectively. Imposing 
the vacuum conditions we can replace $m_1^2$ and $m_2^2$ by $v_1$ and $v_2$.
Then the masses of the heavier bosons ($H$, $H^\pm$ and $A$) 
take the form $m_\Phi^2 = m_3^2/(\sin \beta \cos \beta)^2 + \lambda_i v^2$, 
where $\lambda_i$ is a linear  combination of $\lambda_1$-$\lambda_5$. 
When $m_3^2/(\sin \beta \cos \beta)^2 \gg \lambda_i v^2$, the mass $m_{\Phi}^2$ 
is determined by the soft-breaking scale of the discrete symmetry $m_3^2$, and
is independent of $\lambda_i$. This corresponds to the so-called
decoupling limit. On the contrary, when $M^2$ is limited to be at the weak scale
($M^2 \lsim \lambda_i v^2$) a large value of $m_\Phi^{}$ is realized
by taking $\lambda_i$ to be large; i.e., the strong coupling regime.
However, too large $\lambda_i$ leads to the breakdown of perturbation
theory \cite{Lee:1977yc,Kanemura:1993hm,unitarity2}.
Furthermore, low energy precision data also impose
important constraints on the model parameters \cite{Peskin:1991sw}.
We take into account the following bounds to constrain the 2HDM parameters
\footnote{For comparison, see the recent analysis of Ref.~\cite{osland}.
Even if our results qualitatively agree with those of Ref.~\cite{osland},
our numerical analysis was still necessary to understand whether the specific
scenario studied in the present work is possible within a 2HDM.}:

\begin{enumerate}[i)]
\item Perturbative unitarity \cite{Lee:1977yc},
corresponding to $|a^0(\varphi_A^{}\varphi_B^{}\to\varphi_C^{}\varphi_D^{})|<\xi$
(we take $\xi=1/2$ in our analysis),
where $a^0(\varphi_A^{}\varphi_B^{}\to \varphi_C^{}\varphi_D^{})$
is the S-wave amplitude for the elastic scattering process
$\varphi_A^{}\varphi_B^{}\to \varphi_C^{}\varphi_D^{}$ of the
longitudinally polarized gauge bosons (and Higgs bosons).
These conditions translate into constraints on the 
couplings $\lambda_i$ ($i=1-5$) \cite{Kanemura:1993hm,unitarity2}.

\item Vacuum stability \cite{vacuum-stability}

\item Constraints on oblique-corrections from LEP with the $S$, $T$ and $U$
parameters \cite{Peskin:1991sw}. In particular, the $T$ parameter is such
that $T\simeq\alpha_{EM}^{-1}\Delta\rho$, with $\Delta \rho \leq 10^{-3}$.
The above constraint can be satisfied if a custodial $SU(2)_V^{}$ \cite{CSTHDM}
is approximately conserved and this happens if
(1) $m_{H^\pm}^{} \simeq m_A^{}$, and
(2) $m_{H^\pm}^{} \simeq m_H^{}$ with $\sin^2(\alpha-\beta) \simeq 1$
or $m_{H^\pm}^{} \simeq m_h^{}$ with $\cos^2(\alpha-\beta) \simeq 1$
\cite{Lim:1983re,CSTHDM}

\item B-physics constraints, in particular $\BXsga$ and $B\to\tau\nu$.
Regarding $\BXsga$, the present experimental world average performed 
by HFAG \cite{Barberio:2007cr} is
\beq
\BRga_{\rm exp}=\left (3.55\pm 0.24^{+0.09}_{-0.10}\pm 0.03\right)
\times 10^{-4}
\eeq
while the theoretical estimate performed at the NNLO level
\cite{Misiak:2006ab, Misiak:2006zs} (for the reference value
$E_{\rm cut} = 1.6 \,\rm{GeV}$) is
\beq
\BRga_{\rm SM} = (3.15 \pm 0.23) \times 10^{-4}.
\eeq
The NNLO SM prediction for $\BRga_{\rm SM}$ is lower than $\BRga_{\rm exp}$ by
more than $1 \mysigma$. This fact allows sizable NP contributions with the 
same sign as the SM ones like charged-Higgs boson contributions in 2HDMs.
In the numerics we utilize the formulae presented in Ref.~\cite{Hurth:2003dk}
and updated in Ref.~\cite{Lunghi:2006hc} that take into account the NNLO 
contributions for the SM \cite{Misiak:2006ab, Misiak:2006zs}.

Combining the SM prediction and the experimental results for $\BRga$, we impose 
the constraint
\be
R_{bs\gamma}^{\rm exp} =
\frac{\BR^{\rm exp}(b\to s\gamma)}{\BR^{\rm SM}(b\to s\gamma)}
~=~ 1.13 \pm 0.12~,
\label{eq:bsgamma_exp}
\ee
at the $2\sigma$ level.

Combining the recent $B$-factory results~\cite{:2007xj,Ikado:2006un}~, with the SM
expectation ${\cal B}(\Btaun)^{\rm SM}=G_{F}^{2}m_{B}m_{\tau}^{2} f_{B}^{2}|V_{ub}|^{2}
(1-m_{\tau}^{2}/m_{B}^{2})^{2} /(8\pi \Gamma_B)$,  whose numerical value
suffers from sizable parametrical uncertainties induced by $f_B$ and $V_{ub}$,
it is found that 
\be
R_{B\tau\nu}^{\rm exp} =
\frac{\BR^{\rm exp}(\Btaun)}{\BR^{\rm SM}(\Btaun)}
~=~ 1.07 \pm 0.42~.
\label{eq:Rtn_exp}
\ee
where we have assumed $f_B= 0.216\pm 0.022$ and $V_{ub}=(4.00\pm 0.26)\times1
0^{-3}$ (from the average of inclusive and exclusive semileptonic B decay modes)
by HFAG \cite{Barberio:2007cr}. The decay $\Btaun$ represents a very powerful probe
of the scenario of light charged Higgs \cite{Hou:1992sy,IP,IP2} because it is a
tree-level process and we have that
\beq
 R_{B\tau\nu} = \frac{\BR^{\rm 2HDM}(\Btaun)}{\BR^{\rm SM}(\Btaun)}
=\left[1-\left(\frac{m^{2}_B}{m^{2}_{H^\pm}}\right)~\tan^2\beta
\right]^2~.
 \label{eq:Btn2H}
\eeq

In the case of the decay $B_s\to\mu^+\mu^-$ in a 2HDM we have
$BR(B_s\to\mu^+\mu^-)_{2HDM}\propto\tan^{4}\beta/M^{4}_{H^\pm}$ \cite{Bsmm2HDM}
instead of $BR(B_s\to\mu^+\mu^-)_{SUSY}\propto\tan^{6}\beta/M^{4}_{A^0}$ that
is obtained in SUSY models (see next section).
Therefore $B_s\to\mu^+\mu^-$ does not provide a further constraint on the 2HDM
parameter space once the previous constraints are satisfied.
\end{enumerate}

Applying these constraints we have numerically found the allowed range
for $m_{H^+}$ requiring a pseudoscalar mass $m_A$ in a narrow region,
$150< m_A/\rm{GeV}< 200$, while all the other parameters of the model
including $\tan \beta$ are left free. The upper bound on the $m_{H^+}$
mass (for the imposed range of $m_A$) is found to be $\sim 400$ GeV for any
$\tan\beta$ value by the unitarity and $\Delta \rho$ constraints.
The lower bound on $M_{H^\pm}$ is set by the constraints 
arising from $\BXsga$ ($M_{H^\pm} > 295\,\rm{GeV}$ at $95\%$ 
confidence level independently of $\tan \beta$  
\cite{Misiak:2006zs}). This bound is improved for large $\tan \beta$ values 
($\tan\beta\sim 45\mbox{--}65$) by the $B\to\tau\nu$ constraints. In fact, as
discussed in the next section, the tree-level decay $B\to\tau\nu$ sets a
bound on $\tan \beta/M_{H^+}$ that roughly allows charged-Higgs masses higher 
than 295 GeV  for $\tan \beta=45$ and higher than 420 GeV for $\tan \beta=65$
\footnote{In fact, lighter charged Higgs masses than these values, although
  never lighter than 295 GeV, can be allowed for larger $\tan \beta$ values if
  the SM contribution is canceled by a charged Higgs contribution as
  large as  twice the SM one with opposite sign.}.
Therefore a generic 2HDM of type II can still be compatible with a range 
between 150 and 200 GeV for the mass of the light pseudoscalar Higgs boson,
although the charged Higgs is always constrained to be above 295 GeV by 
$\BXsga$.

However, we have to recall that it is very difficult to accommodate the present
discrepancy for the muon anomalous magnetic moment in a 2HDM scenario.

Notice that a scenario with a light pseudoscalar Higgs boson with mass
$M_A\leq 200 \rm{GeV}$ and a charged Higgs with mass $M_{H^\pm}\geq 300\rm{GeV}$
is not compatible with minimal SUSY frameworks.

\section{Supersymmetric models}
\label{sec:MSSM}

The Higgs sector of the MSSM is a special case of Eq.~(\ref{pot}).
The MSSM has been extensively studied in the literature (see
e.g. \cite{MSSM,Chung:2003fi} and references therein)
and the presence of the different supersymmetric partners of
the SM particles increases the phenomenological constraints to satisfy \cite{fcncreview}.
Therefore, the first question we have to answer is whether it is possible or
not  to obtain a pseudoscalar Higgs boson of a mass below 200 GeV in the MSSM 
satisfying simultaneously all the different constraints.
We will answer this question basically in two versions of the MSSM: the
Constrained MSSM (CMSSM) and the non-universal Higgs mass (NUHM) MSSM
\cite{Olechowski:1994gm,Berezinsky:1995cj,Nath:1997qm,Bartl:2001wc,Ellis:2002wv,Ellis:2002iu,Cerdeno:2004zj,Baer:2005bu}.
Then we comment about other SUSY-breaking mediation mechanisms such as
gauge mediation and anomaly mediation models.

\subsection{CMSSM and NUHM}

The CMSSM is fixed by 4 initial parameters : $m_0$, $M_{1/2}$, $A_0$ and $\tan \beta$ 
plus the sign of the $\mu$ parameter. However,  the sign of the $\mu$ parameter is bound 
to be positive by the requirement of a correct prediction to the muon anomalous
magnetic moment and the $B \to X_s \gamma$ branching ratio.
Before considering in detail the different indirect constraints it is useful
to identify the possible regions of CMSSM parameter space that can accommodate 
a pseudoscalar mass below 200 GeV. 

In order to establish the $m^2_{H^+}$, $m^2_{A}$ dependence on the input
parameters of the CMSSM, let us consider the following tree-level
expressions and their approximate values at medium to large values of 
$\tan\beta$:
\begin{eqnarray}
\mu^2& =& \Frac{m_{H_d}^2(m_t) - m_{H_u}^2(m_t)  \tan^2 \beta}{\tan^2 \beta -1}
 -\frac{M_Z^2}{2} \simeq - m_{H_u}^2(m_t)  -\frac{M_Z^2}{2} \label{TLmu}\\
m_A^2 &=&  m_{H_d}^2(m_t) + m_{H_u}^2(m_t) + 2 \mu^2 \simeq  m_{H_d}^2(m_t) - m_{H_u}^2(m_t) - M_Z^2 \label{TLma}
\end{eqnarray} 
The mass of the charged Higgs and the pseudoscalar\footnote{Here we are
  considering a real MSSM with zero phases in the $\mu$ and trilinear
  parameters.} are very similar in the MSSM, as at tree-level they satisfy
the relation: 
\bea
m^2_{H^+} = m^2_{A} + m_W^2\, .
\label{relTL}
\eea
These masses are then obtained from the electroweak values $m_{H_u}^2(m_t)$ and 
$m_{H_d}^2(m_t)$ which are determined through the RGEs
\bea
16\pi^2\frac{d}{dt}m^{2}_{H_u} &\simeq& 6 X_t - 6 g_2^2 M^2_{1/2}\\
16\pi^2\frac{d}{dt}m^{2}_{H_d} &\simeq& 6 X_b + 2X_{\tau}- 6 g_2^2 M^2_{1/2}
\label{RGEmH}
\eea
with $X_f = y^2_f\left( m^{2}_{H} + m^{2}_{\tilde{f}_L} +
m^{2}_{\tilde{f}_R}+ A^2_{f}\right)$ ($f=t,b,\tau$).
As it is well known, $m^{2}_{H_u}(m_t) < 0$ in the entire SUSY parameter space 
due to the large RGE effects proportional to $y^2_t$.
The approximate numerical solution for $m^{2}_{H_u}(m_t)$, valid for any $\tan
\beta$ value, is \cite{Bartl:2001wc}
\beq
m^{2}_{H_u}(m_t)\simeq -0.12~ m^2_0 - 2.7~ M^{2}_{1/2} +0.4~
A_{0}M_{1/2}-0.1~ A^{2}_{0} \, ,
\label{mHu}
\eeq
clearly showing that $m^{2}_{H_u}(m_t)<0$. 
On the contrary, $m^{2}_{H_d}(m_t)$ crucially depends on $\tan\beta$. 
For instance, if we assume low/moderate $\tan\beta$, i.e. $\tan\beta\leq 10$, 
we can neglect to first approximation $X_{b,\tau}$ in Eq.~(\ref{RGEmH}) and
the LO solution for $m^{2}_{H_d}(m_t)$ is
$m^{2}_{H_d}(m_t)\simeq m^2_0 + 0.5~M^{2}_{1/2}$. In this regime, 
$m^{2}_{H_d}(m_t)>0$ and thus both $m^{2}_{H_d}(m_t)$ and $m^{2}_{H_u}(m_t)$ 
provide positive contributions to $m_A^2$ in Eq.~(\ref{TLma}).
For larger $\tan\beta$ values, negative RGE effects proportional to  
$y^{2}_{b,\tau}$ reduce $m^{2}_{H_d}(m_t)$  until the limit case where 
$m^{2}_{H_d}\simeq m^{2}_{H_u}<0$ when $y^2_b\sim y^2_t$. In this large 
$\tan\beta$ regime, 
$m^{2}_{H_d}$ and $m^{2}_{H_u}$ provide opposite contributions to 
$m_A^2$ in \eq{TLma} that, indeed, can result strongly reduced. 
As we will see below, our numerical analysis confirms that we find light
 pseudoscalar masses, $m_A \leq 200$ GeV, only for $\tan \beta >50$.  

At large $\tan \beta$, the requirement of correct EW symmetry breaking and 
a neutral LSP sets important constraints on the allowed 
 $(m_0,M_{1/2},\tan\beta)$ values. In particular in CMSSM models, 
the lightest stau mass is $M^{2}_{\tilde{\tau}_1} \sim M^{2}_{\tilde{\tau}_R} - 
 m_{\tau}\mu\tan\beta$  where at leading log, $M^{2}_{\tilde{\tau}_R}\simeq  
m_{0}^{2} -0.36(3 m_0^2+A^{2}_{0})y^{2}_{\tau}] +0.13 M^{2}_{1/2}$ and  
 $y_{\tau}=m_{\tau}\tan\beta\sqrt{2}/v$ (with $v = 246$ GeV). 
Thus, the LSP condition $M_{\tilde{\tau}_1} > M_{\chi^{0}_{1}}$ generates
 a lower bound for $m_0$ that increases with increasing $\tan\beta$.
For instance, at $\tan \beta =50$ the minimum value of $m_0$ is 200 GeV 
 for $m_{1/2} = 200$ GeV. However, the $m_0^2$ dependence of $m_A^2$ is 
quite mild, as a result of the large negative RGE effects driven by 
 $y^2_t,y^2_b\sim 1$. So, at large $\tan\beta$, the $m_A^2$ mass is almost 
determined by the $M_{1/2}^2$ contribution while a relatively large $m_0$
affects the Higgs mass only marginally. On the other hand,  
the allowed values for the trilinear parameter $A_0$ are also constrained by 
the requirement of absence of charge and colour breaking minima and it turns   
out typically that $-3<A_0/m_{0}<3$. In the analysis of the flavour physics 
observables the value of  $A_t$ (together with the value of $\mu$) plays a  
particularly important role. The RGE equation for $A_{t}$ is: 
\beq   
16\pi^2\frac{d}{dt}A_{t} \simeq 12 A_{t} y^2_t - \frac{32}{3}g_3^2 M_3\,,
\label{RGEAt} 
\eeq 
where $t=\log(Q/M_{\rm GUT})$. The LO solution of \eq{RGEAt} provides the  
approximate result $A_{t}(m_t)\simeq 0.25 A_{t}(0) - 2 M_{1/2}$. 
A relevant observation for the following discussions is that it is always
possible to get small $|A_{t}(m_t)|$ values by opportunely selecting  
$A_{t}(0)$ and $M_{1/2}$. 
In summary the above qualitative considerations clearly show that we can  
have relatively small heavy Higgs masses in the CMSSM with  
large $\tan \beta$, small values of $M_{1/2}$ and relatively large values of     
$m_0$. This is the region of CMSSM parameter space that we will explore numerically in  
detail below.

Similarly, the NUHM MSSM is a simple extension of the CMSSM where the initial
values of the Higgs masses, $m_{H_{d,0}}$ and  $m_{H_{u,0}}$ are different 
from the rest of the sfermion masses, $m_0$ at the mediation
scale \cite{Olechowski:1994gm,Berezinsky:1995cj,Nath:1997qm,Bartl:2001wc,Ellis:2002wv,Ellis:2002iu,Cerdeno:2004zj,Baer:2005bu,Carena:2007aq,Carena:2006ai}.  
In this model we can expect that the additional freedom of varying
the initial values  of $m_{H_{d,0}}$ and $m_{H_{u,0}}$ can help to reduce
$m_A^2$. In NUHM models, \eq{mHu} is changed to \cite{Bartl:2001wc} 
\beq
m^{2}_{H_u}(m_t)\simeq -
0.75~ m^2_0 +0.63~ m_{H_{u,0}}^2 - 2.7~ M^{2}_{1/2} +0.4~ 
A_{0}M_{1/2}-0.1~ A^{2}_{0}\, .
\eeq 
Similarly to the CMSSM case, the value of 
$m^{2}_{H_d}(m_t)$ depends strongly on $\tan\beta$. However, given that the  
bottom and tau Yukawa couplings are smaller than the top Yukawa up to values
of $\tan \beta \gsim 50$, it is clear that we can expect the coefficient of  
$m_{H_{d,0}}$ to be positive. Then replacing these values in \eq{TLma}, we can
see that the contribution of $m_{H_{u,0}}$ to $m_A^2$ is negative while the  
contribution of $m_{H_{d,0}}$ is positive (see for instance Table III in 
\cite{Bartl:2001wc}). Therefore we can  reduce $m_A^2$ in the NUHM with respect
to the CMSSM for $m_{H_d} < m_{H_u}$.  

\begin{figure}  
\hspace{-0.3 cm}
\includegraphics[scale=0.40]{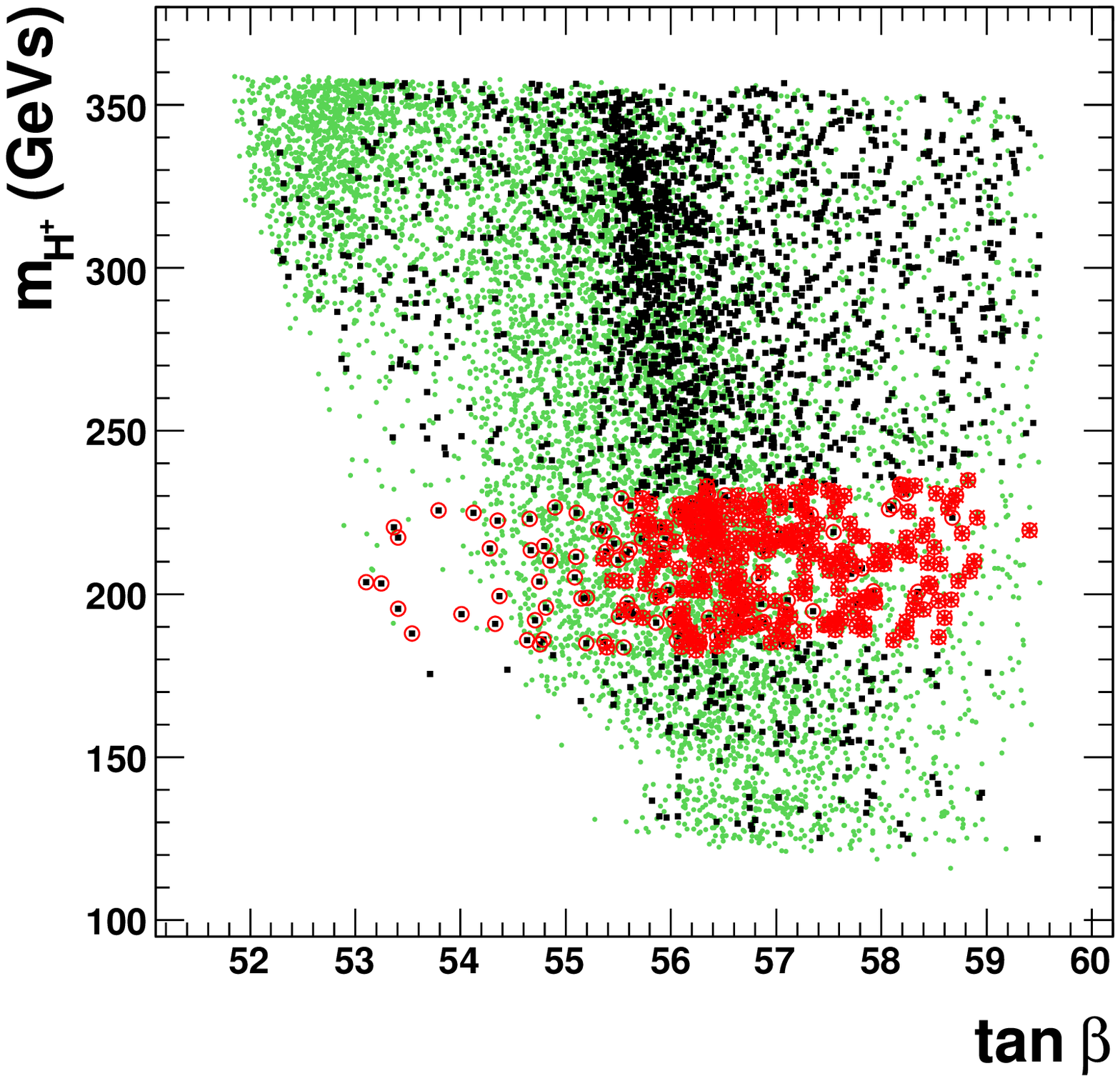}    
\hspace{-0.3 cm}   
\includegraphics[scale=0.439]{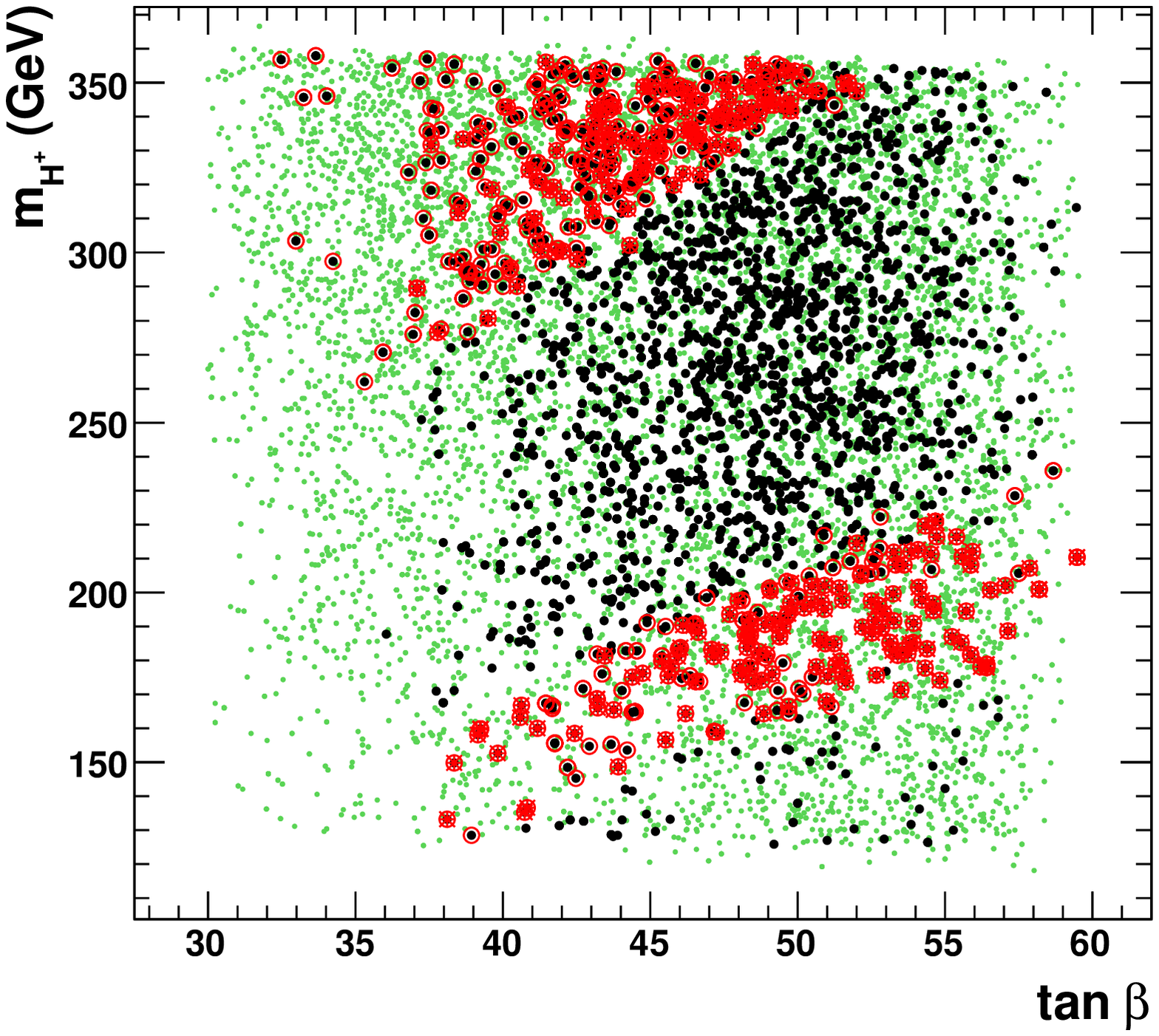} 
\hspace{-0.3 cm}   
\caption{Light values of $m_{H^+}$ as a function  of
$\tan \beta$. In the CMSSM (left) we scan on $M_{1/2}\leq 600$ GeV,   
$900 <m_0<2500$ GeV, $\tan \beta > 30$ and $-3 m_0<A_0< 3 m_0$.  In the NUHM 
(right), in addition we allow $0.75 < m_{H_u}/m_0,  m_{H_d}/m_0  <  1.25$. 
Green (light grey) points satisfy all direct bounds  on scalar 
and gaugino masses. Black points satisfy also the main indirect  constraints
and $\Omega_{\chi} < 0.14$ as explained in the text and red circles are
points that  in addition predict a BR($B \to \tau \nu$) within the 
experimental range. Crossed red circles satisfy  $0.08< \Omega_{\chi} <
0.14$.}  
\label{fig:mH+tgb}  
\end{figure}
\begin{figure} 
\includegraphics[scale=0.50]{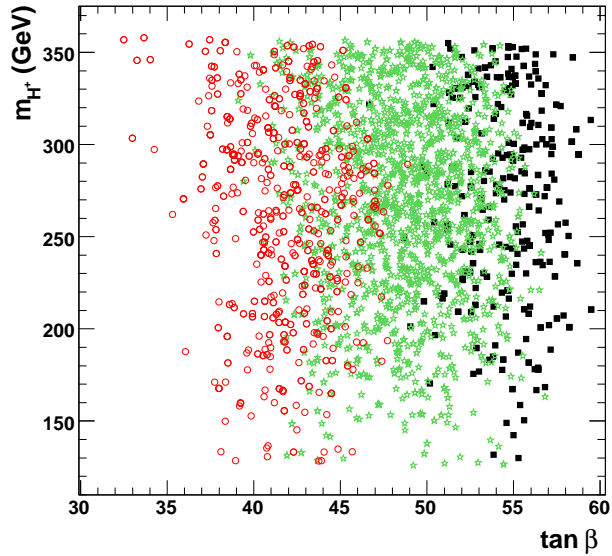}    
\caption{Values  of $m_{H^+}$ as a function of
$\tan \beta$ in the presence of non-universality at $M_{\rm GUT}$. Black 
squares correspond  to points where $( m_{H_u}-m_{H_d})/m_0 \leq 0$,   
green (light grey) stars correspond to points with  $0\leq(
m_{H_u}-m_{H_d})/m_0  \leq 0.3$ and red open circles to points where   
$0.3\leq( m_{H_u}-m_{H_d})/m_0$. All these points satisfy all direct and
indirect bounds with the  exception of BR($B \to \tau \nu$). The upper  bound on
the dark matter abundance is also imposed.} 
\label{fig:mHNUHM}  
\end{figure}

   
 Our numerical analysis below is done using RGE at two loop order 
\cite{Martin:1993zk} taking into account the complete flavour structure 
and the masses calculated at one-loop order \cite{Pierce:1996zz} using 
SPheno \cite{Porod:2003um}. 
In particular for the $\mu$ parameter and the neutral Higgs masses two-loop
corrections are added \cite{Degrassi:2001yf}. We always impose the direct
constraints on sfermion, gaugino and chargino masses from LEP and
Tevatron \cite{Yao:2006px}. 
In Figure \ref{fig:mH+tgb}  we show the values of $m_{H^+}$ as  a function of
$\tan \beta$ for $\tan \beta>30$ in a scatter plot with    
$M_{1/2}\leq 600$ GeV and  $900 <m_0<2500$ GeV in the CMSSM and NUHM. 
All the points in these  figures, including green points satisfy  all direct 
bounds on scalar and   
gaugino masses. Ignoring indirect constraints for the moment, the  most 
interesting feature here  is the strong dependence of the mass with 
$\tan \beta$. We see  that, indeed as discussed above, we can obtain  charged 
Higgs masses below $200$ GeV in the CMSSM only for very large values of  
$\tan \beta$, $\tan \beta \geq 53$.  Therefore, before imposing the dark
matter  and indirect constraints,  it is   
possible to obtain charged Higgs masses below 200 GeV in the CMSSM for 
$\mu >0$, $\tan \beta \geq 54$,    $m_0 \geq 900$ GeV and $M_{1/2}\leq 400$ 
GeV. In the NUHM case, we have allowed a small departure from universality 
for the Higgs masses that can be 25\% lighter or
heavier that the common soft-mass   $m_0$ at the GUT scale. 
In Figure \ref{fig:mHNUHM} we can see the effect of a small
non-universality in the GUT Higgs   masses in the mass of the charged Higgs at
$M_W$.  
As expected, the difference  in the initial values of  
Higgs masses has a strong impact on the mass of the charged 
Higgs although we allow  only a 25\% departure for $m_{H_u}$ and 
$m_{H_d}$ from  $m_0$. This becomes more pronounced if we allow for a larger
breaking of universality.
In the right-hand side plot of figure \ref{fig:mH+tgb} we see that  this small departure
from universality is enough  to obtain charged Higgs 
masses smaller than 200 GeV for values of $\tan \beta$ as  low as 30 before
imposing the indirect constraints.

Next, we must check that in this region of the parameter space it
is possible to satisfy all the indirect constraints, specially those
arising from processes enhanced by powers of $\tan\beta$, namely
$BR(B\to\tau\nu)$, $BR(B\to X_s\gamma)$, $B_s\to\mu^+\mu^-$ and the muon
anomalous magnetic moment $(g-2)$.
The behaviour of these observables in the large $\tan\beta$ regime was
discussed in detail in
Refs.~\cite{IP,IP2,Lunghi:2006uf,ellisfirst,Hamzaoui:1998nu,Choudhury:1998ze,Babu:1999hn,Chankowski:2000ng,Isidori:2001fv,Bobeth:2002ch,Buras:2002wq,Buras:2002vd,Kane:2003wg,Dedes:2004yc},
\footnote{However, these analyses do not include the Higgs mediated FCNC contributions 
pointed out in Ref.~\cite{Freitas:2007dp}. As shown in Ref.~\cite{Freitas:2007dp}, the
renormalization of both $\tan\beta$ and the Higgs masses may lead to sizable effects
for $\Delta M_{B_{s,d}}$ in the narrow region where $M_{A}\lsim 160$~GeV.
Although this is exactly the region relevant for our analysis, we have checked numerically
that the inclusion of these new effects do not lead to any further constraints on our scenario.}
and here we only summarize the main features.

On general grounds, the simultaneous requirement of a light $M_A$ and large $\tan\beta$
values strongly enhances Higgs mediated FCNC effects. Thus, very special conditions are
necessary in order to satisfy all the phenomenological constraints, with special attention
to the $B_s\to\mu^+\mu^-$ decay.

The decay $B\to X_s\gamma$ receives the dominant contributions from the W-boson,
the charged Higgs and the chargino diagrams. Gluino and neutralino
contributions depend on radiatively generated Mass Insertions through the RG 
evolution, of order  $c~ V_{tb} V_{ts}$, with $c$ a loop factor. 
Hence gluino and neutralino are subdominant with respect to the previous ones 
that do not have this additional loop suppression. However all contributions 
at one-loop order have been included in our numerical analysis below.

In the numerical analysis, we have imposed the allowed range for 
$BR(B\to X_s\gamma)$ as reported in Eq.~(\ref{eq:bsgamma_exp}) and we
have evaluated $BR(B\to X_s\gamma)$ including the SM effects at the NNLO
and the NP contributions at the LO.
The charged Higgs contribution has always the same sign as the SM contribution.
This contribution depends basically on the charged Higgs mass and it
depends mildly on $\tan\beta$ through the threshold corrections to the
bottom Yukawa coupling. Even though in the MSSM these threshold corrections
reduce the size of the charged Higgs contribution (for $\mu>0$)
compared to the 2HDM case, a charged Higgs of about 200 GeV already saturates
the allowed range for $BR(B\to X_s\gamma)$. Therefore, there is  no space left
for a chargino contribution with the same sign as the SM one.
The relative sign between the chargino and the SM amplitudes is given
by $\hbox{sign}(A_t~\mu)$. In the MSSM, except for very large $A_0 > 8
M_{1/2}$, we have always that $A_t(M_W)<0$.
Under these conditions, a $M_A\leq 200 \rm{GeV}$ forces to choose $\mu>0$
in order to get the necessary destructive  interference between charged
Higgs and chargino contributions to $B\to X_s\gamma$.
Moreover, the chargino amplitude, $A_{\tilde{\chi}^-}$, is  proportional to 
$A_{\tilde{\chi}^-}\propto
[\mu A_t/m^{4}_{\tilde{q}}]\times \tan\beta$. Thus to keep
$A_{\tilde{\chi}^-}$ under control with very large values  
of $\tan\beta$, we need large sfermion masses and small $A_t$ and $\mu$.
In fact this is precisely the situation we find in our numerical analysis,
where all the squark masses are above one TeV and $\mu$, related to $m_{A}$,
is relatively small.

The value of $A_0$ is scanned in the region $-3<A_0/m_{0}<3$ while we consider
$M_{1/2} \leq 600$ GeV.
If we remember that $A_t(m_t)\simeq 0.25 A_0 -2M_{1/2}$, it is relatively
easy to find small values for $A_t(m_t)$ when $A_0 >0$ and for large $m_0$
compared to $M_{1/2}$, as it happens in the scenario we are considering.

Likewise, it is easy to find the SUSY contributions to the muon anomalous
magnetic moment of the required size to explain its discrepancy with the
SM expectation $\Delta a_{\mu} = a_{\mu}^{\rm exp}  - a_{\mu}^{\rm SM}
\approx(3 \pm 1)\times 10^{-9}$
\cite{Lunghi:2006uf,Dedes:2001fv,Baek:2002wm}.
This discrepancy can be accommodated only with a positive $\mu$ sign, in agreement
with the $b\to s \gamma$ requirements.
The main SUSY contribution to $a^{\rm MSSM}_\mu$ is  provided by the loop exchange
of charginos and sneutrinos. The basic features of the supersymmetric contribution
to $a_\mu$ are correctly reproduced by the following approximate expression:
\beq
\frac{a^{\rm MSSM}_\mu}{ 1 \times 10^{-9}}
\approx    1.5\left(\frac{\tan\beta }{50} \right)
\left( \frac{1000~\rm GeV}{m_{\tilde \nu}} \right)^2~,
\label{eq:g_2}   
\eeq
which provides    a good approximation to the full one-loop result
\cite{Bartl:2003ju} when the chargino masses are substantially
lighter then the slepton masses, as it happens in our case. 

%
%
%
The SUSY contributions to ${\rm BR}(B_s\to\mu^+\mu^-)$  can be summarized
by the following approximate formula 
\beq 
{\rm BR}(B_s\to\mu^+\mu^-)\simeq   
\frac{4\times10^{-8}}{\left[1+0.5 \times\frac{\tan \beta}{50}\right]^4}
\Bigg[\frac{\tan\beta}{50}\Bigg]^6  
\left(\frac{160 \rm{GeV}}{M_A}\right)^4 
\left(\frac{\epsilon_{Y}}{4\times 10^{-4}}\right)^{2}  
\label{bsmumuSUSY} 
\eeq 
where $\epsilon_{Y}$ is defined through the flavor violating Yukawa interactions
\beq 
\CL_{A}= \Frac{ig_2}{2M_W}~m_b~ \Frac{\epsilon_{Y}V_{ts}\tan^{2}_{\beta}}
{(1+\epsilon_{0}\tan_{\beta})^2}~ \bar{b}_R s_L A + h.c.\, ,
\eeq 
It receives contributions both from charginos and gluinos\footnote{Notice even though $\left(\delta_{LL}^d\right)_{23}/ V_{ts} = O(0.1)$, gluino contributions to the effective $H_u b s$ vertex do not decouple and these contributions can play and important role when chargino contributions are reduced through a small $A_t$.} thus, we can write
$\epsilon_{Y}=\epsilon^{\tilde{\chi}^-}_{Y}+\epsilon^{\tilde{g}}_{Y}$ with
\beq
 \epsilon^{\tilde{\chi}^-}_{Y} \simeq
-  \frac{1}{16\pi^2}\frac{A_t}{\mu} H_2(y_{u_R},y_{u_L})
\,,\qquad 
  \epsilon^{\tilde{g}}_{Y}\simeq-
 \frac{2\alpha_3}{3\pi}\frac{\mu}{M_{\tilde{g}}}H_3(x_{d_R},x_{d_R},x_{d_L})
  \frac{(\delta_{LL}^d)_{23}}{V_{ts}}\,, 
\label{epsu1}
  \eeq 
 with $y_{q_{R,L}}=M^2_{\tilde{q}_{L,R}}/|\mu|^2$,
$x_{q_{R,L}}=M^2_{\tilde{q}_{L,R}}/M^{2}_{\tilde{g}}$, $(\delta_{LL}^d)_{23}$
the left-handed squark mass insertion at the electroweak scale  
and the loop functions are such that $H_2(1,1)=-1/2$, $H_3(1,1,1)=1/6$.

From \eq{bsmumuSUSY}, if $M_A=160~\rm{GeV}$ and $\tan\beta=50$,
we can saturate the present experimental upper bound on $BR(B_s\to\mu^+\mu^-)$
when $\epsilon_{Y}\simeq 4\times 10^{-4}$. On the other hand, in the limit of
all the SUSY masses and $A_t$ equal, the pure chargino contribution
$\epsilon^{\tilde{\chi}^-}_{Y} \sim 3\times 10^{-3}$.
However, as discussed above, $A_t<0$ in most of the parameter space 
of SUGRA models so that, irrespective to the $\mu$ sign,
sgn$(\epsilon^{\tilde{g}}_{Y}/\epsilon^{\tilde{\chi}^-}_{Y})=-1$ 
is unambiguously predicted when the sign of $(\delta_{LL}^d)_{23}$ is
negative. We remind the reader that, even assuming a 
flavor blind soft sector at the GUT scale, running effects from $M_{GUT}$ 
down to the electroweak scale generate off-diagonal entries in the squark 
mass matrix as both type of Yukawa couplings, $Y_u$ and $Y_d$, contribute.
Then we have $(\delta_{LL}^d)_{23}\simeq c \times V_{ts}$ with $c <0$ and 
typically $O(0.1)$ although it can be even close to 1 in special regions. 
Therefore cancellations between chargino and gluino contributions to 
$\epsilon_{Y}$ can be important when both contributions have similar sizes.
$BR(B_s\to\mu\mu)$ is also reduced when both $\epsilon^{\tilde{g}}_{Y}$ and 
$\epsilon^{\tilde{\chi}^-}_{Y}$ are small, i.e.  when  
$(\delta_{LL}^d)_{23}$ and $A_t$ respectively are small.  
Given that $A_t\simeq 0.15 ~A_0-2~M_{1/2}$ (where the $M_{1/2}$ contributions
are RGE induced and $-3\leq A_0/m_0 \leq 3$), it is clear that we can lower
$A_t$  for large and positive $A_0$ values and moderate/small $M_{1/2}$; in this
same region, the coefficient $c$ in $(\delta_{LL}^d)_{23}$ is large, 
and this makes the cancellation mechanism more effective. In the region of
parameter space we explore, this is exactly the situation: we have both $m_0$
and $A_0$ large and $M_{1/2}$ small. Then $A_t$ (and also $\mu$) are relatively
small when compared to the heavy sfermion masses. The arguments of the loop
functions, specially  $y_{q_{R,L}}$, are large and then $H_2(x>>1,y=x)\simeq
-1/x$ also reducing the chargino contribution.

Finally let us consider the $B\to\tau\nu$ decay. As shown in \eq{eq:Rtn_exp},
the ratio between the experimentally measured branching ratio and the SM
expectation is given by $R_{B\tau\nu}^{\rm exp} = 1.07 \pm 0.42$.
On the other hand, in SUSY, the charged-Higgs exchange contribution is
\cite{Hou:1992sy,IP,IP2,AR}
\beq
 R_{B\tau\nu} = \frac{\BR^{\rm SUSY}(\Btaun)}{\BR^{\rm SM}(\Btaun)}
=\left[1-\left(\frac{m^{2}_B}{m^{2}_{H^\pm}}\right)
 \frac{\tan^2\beta}{(1+\epsilon_0\tan\beta)}
\right]^2~,
 \label{eq:Btn}
\eeq
where non-holomorphic corrections to the down-type Yukawa coupling
have been included.
As evident from Eq.~(\ref{eq:Btn}), $\Btaun$ represents a very powerful
probe of the scenario we are exploring \cite{Hou:1992sy,IP,IP2}.
In contrast to $B_s\to\mu^+\mu^-$, $\Btaun$ is a tree-level process,
thus, in this last case, there is no way to reduce the size of the NP
contribution when $\tan\beta$ is large and the heavy Higgs is light.
In fact, as we show below, we find that charged Higgs masses below 200 GeV
are only possible when the observed branching ration is obtained through a
SUSY contribution twice the SM one (with opposite sign).

Similarly, charged scalar currents mediated by the charged Higgs affect also
the process $K \to l \nu$ with $l=e,\mu$. The new physics effect in the ratio
$R_{K\mu \nu} = B^{\rm SUSY}(K\to \mu \nu)/ B^{\rm SM}(K\to \mu \nu)$ would
be obtained from \eq{eq:Btn} with the replacement $m_B^2 \to m_K^2$. Although
the charged Higgs contributions are now suppressed by a factor $m_K^2/
m_B^2\simeq 1/100$, this is well compensated by the excellent experimental
resolution \cite{Ambrosino:2005fw,flavianet} and the good theoretical control.
However, given that these new physics effects are at the \% level, we would need
a theoretical prediction for the SM contribution at the same level to use this
decay as an effective constraint. We would then need an independent determination
both of $f_K$ (possibly from lattice QCD) and $V_{us}$. At present unquenched
lattice calculations of $f_K$ are not well established and precise enough.
The above argument for $K\to l\nu$ does not apply to $B\to\ell\nu$.
In fact, even if the $f_B$ and $V_{ub}$ uncertainties are much larger that the
$f_K$ and $V_{us}$ ones, they cannot hide in any way the huge NP effects in
$B\to\ell\nu$ arising in our scenario.
Therefore, although it may play an important role in the future,
we do not include the constraints from $K\to l\nu$ in the following.

In Fig. \ref{fig:mH+tgb}  we can see the effect of these indirect constraints.
Here, green (light grey) points satisfy only direct bounds while black points 
satisfy also the constraints from ${\rm BR}(B\to X_s\gamma)$,
${\rm BR}(B_s\to\mu^+\mu^-)$, the anomalous magnetic moment of the muon
$(g-2)_\mu$ and the upper limit on the dark matter abundance $\Omega_{\chi}<0.14$.
Both in the CMSSM and the NUHM we see that it is rather easy to satisfy these 
indirect constraints due to the relatively heavy sfermions and small $\mu$ and $A_t$.
As expected, the main constraint here comes from the process $\Btaun$ that corresponds
to the red circles in this figure. In the case of the CMSSM, we are bound to values
of $\tan\beta>53$ for $m_{H^+}<200$ GeV.
Therefore, $\Btaun$ sets the lowest allowed value of $m_{H^+}$ to 180 GeV. We have to emphasize again the importance of this tree-level constraint in this scenario. In the absence of this constraint, all black points would be allowed and hence charged Higgs masses as low as 120 GeV would be possible in the CMSSM.

In the case of the NUHM, smaller values of $\tan \beta$ can still produce light charged Higgs masses as seen in Fig. \ref{fig:mHNUHM}. In the plot on the right-hand side of Fig.
\ref{fig:mH+tgb} we can clearly see the dependence of this decay on  $\tan \beta/m_{H^+}$. Here it is possible to obtain charged Higgs masses as low as 120 GeV for $\tan \beta
= 40$. Also in this plot we can see that without this cancellation of the SM contribution, the lower bound on the charged Higgs mass would be 350 GeV for values of $\tan \beta =50$. That is why these points are not visible in the CMSSM plot.

\begin{figure}   
\hspace{-0.3 cm}  
\includegraphics[scale=0.40]{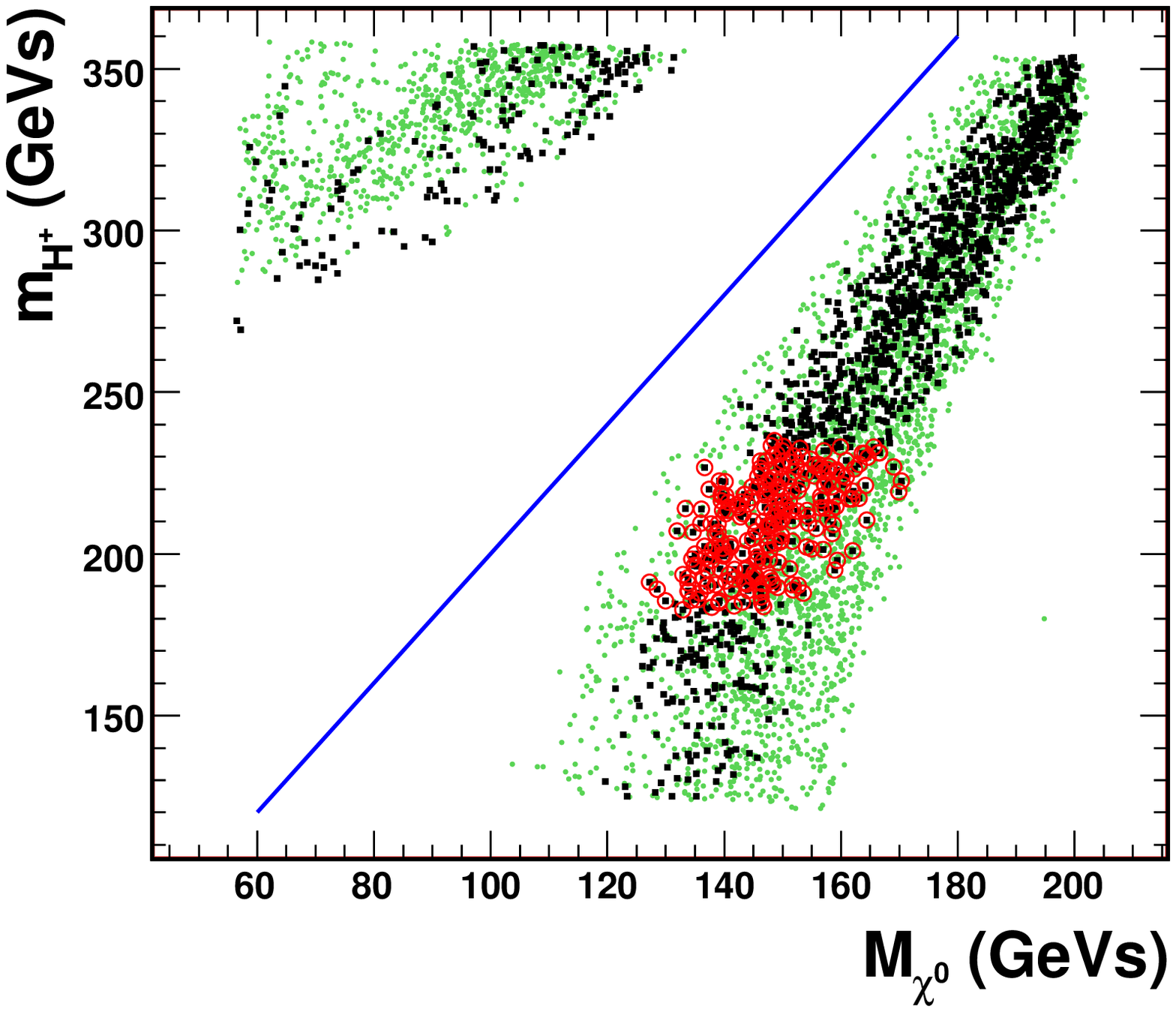}  
\hspace{-0.3 cm}  
\includegraphics[scale=0.40]{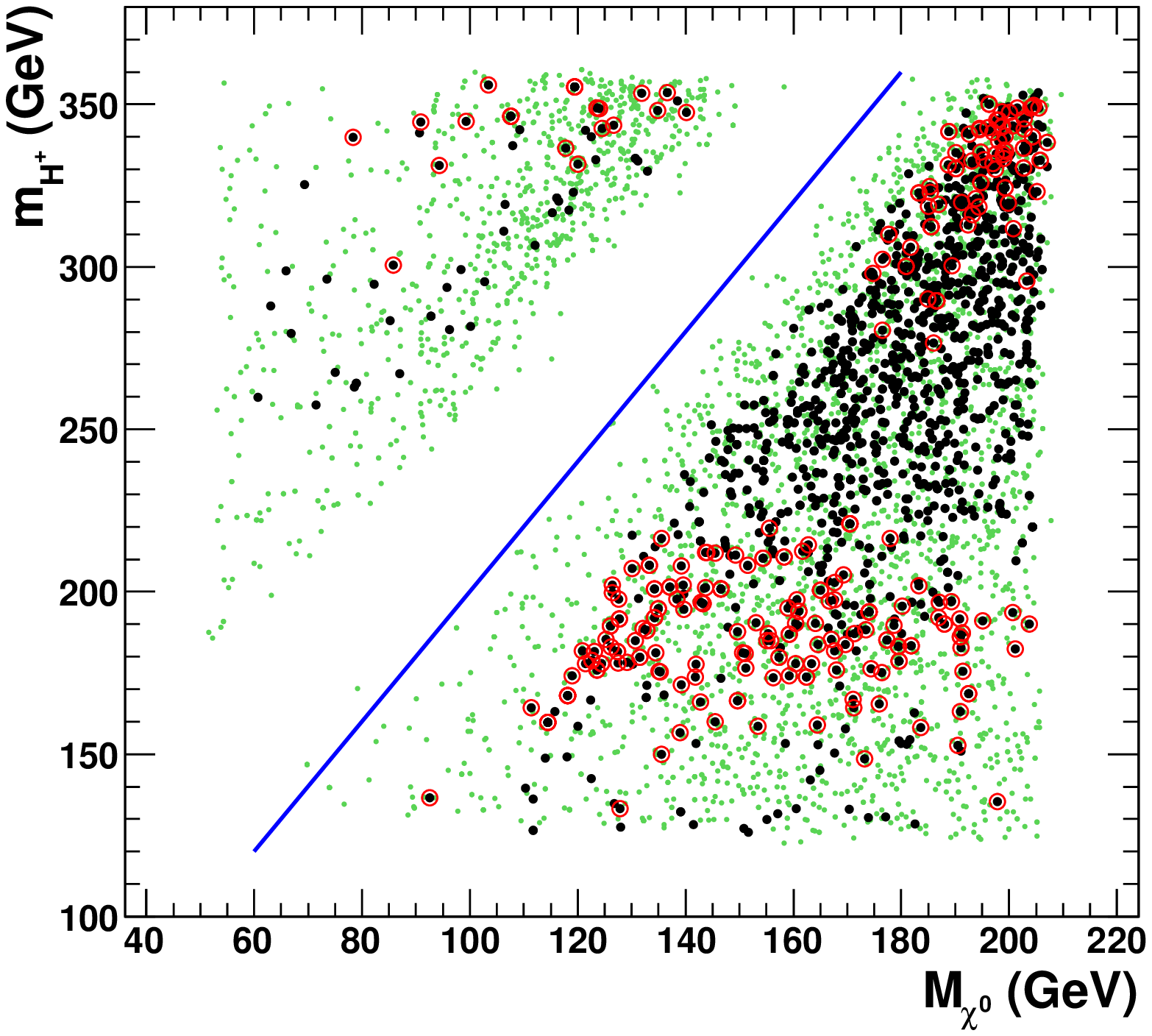}  
\hspace{-0.3 cm}  
\caption{Charged Higgs and lightest neutralino mass for the same points as  
  Figure \ref{fig:mH+tgb} both in the CMSSM (left) and in the NUHM (right). 
Green (light grey) points satisfy all direct bounds  on scalar   
and gaugino masses and the dark matter constraint $0.08<\Omega_{\chi} <
0.14$. Black points satisfy also the main indirect constraints  
 as explained in the text. Red circles are
points that  in addition predict a BR($B \to \tau \nu$) within the 
experimental range.}  
\label{fig:lHfunnCMSSM} 
\end{figure}   

Finally in  Figure \ref{fig:lHfunnCMSSM} we analyze the dark matter constraints
on these points. The requirement of a correct dark matter abundance sets important  
restrictions on the allowed parameter space
 \cite{Berezinsky:1995cj,Nath:1997qm,Cerdeno:2004zj,Baer:2005bu,Dermisek:2003vn,Baek:2004et,Dermisek:2005sw}. However, in this region of large $m_0$ and small $M_{1/2}$ this constraint 
is relatively easy to satisfy.  We can see that both in the CMSSM and the NUHM all our 
points cluster in the funnel region on both sides of the line $m_{H^+} = 2 m_\chi$.
In fact in the  NUHM we can see that there are points (not allowed by indirect
constraints but satisfying the dark matter bound) much closer than expected
to this central  line. This is possible due to the fact that
the annihilation cross section is proportional to $\tan^2 \beta$ and for lower
values of $\tan\beta$ the allowed region is much closer to the resonance.
Therefore we conclude that these points correspond basically to the funnel
region, although significant contributions from other annihilation processes
also occur.

To conclude this section, we would like to comment on the differences
between our analysis and the analysis presented in Ref.~\cite{Ellis:2007ss}.
While on general grounds we agree with the conclusion of Ref.~\cite{Ellis:2007ss},
there are some relevant differences that we would like to emphasize.
First, the authors of Ref.~\cite{Ellis:2007ss} state that it is not possible
in the CMSSM to find a light pseudoscalar reproducing the small CDF
discrepancy consistently with all the constraints.
In contrast, we have shown that a CMSSM can still accommodate such a scenario.
Moreover, the authors of Ref.~\cite{Ellis:2007ss} also claim that
there exists a lower bound for $BR(B_s\to\mu\mu)$ of $BR(B_s\to\mu\mu)>2\times 10^{-8}$
in the region compatible with all the constraints.
We disagree with this statement and we find no lower bound for $BR(B_s\to\mu\mu)$.
Finally, in contrast to Ref.~\cite{Ellis:2007ss}, we emphasize the fundamental
impact of $B\to\tau^+\nu^-$. As we have discussed in detail, $B\to\tau^+\nu^-$
represents probably the most important constraint/probe of this scenario.

 \subsection{Other mediation mechanisms}

In gauge-mediated SUSY breaking scenarios~\cite{Giudice:1998bp}, the SUSY 
breaking is transmitted to the MSSM sector through gauge interactions. In the 
minimal gauge-mediation (MGM) model, the messenger 
fields get fermionic masses $\Mess{} = \lambda
\vevS$ and scalar masses $m^2 = |\lambda \vevS|^2 \pm
|\lambda \vevFS|$ through their Yukawa couplings to a singlet field $S$.  
Supersymmetry breaking is then transmitted to the MSSM gauginos and scalars 
through one-loop and  two-loop diagrams respectively and we obtain:
\bea
\Mbar{a} &=&  N~{\alphbar{a}\over 4\pi} \Lambda g(x) \equiv \Mhat{a}
g(x)\, ,\label{eq:gmasses}\\  
\mbar{\widetilde{\alpha}}^2 &=&
{2  \Lambda^2 ~ N} \left[C_3 \left({\alphbar3\over 4\pi}\right)^2 +
                  C_2 \left({\alphbar2\over 4\pi}\right)^2 +
                   {3\over5} Y^2
\left({\alphbar1\over 4\pi}\right)^2\right] f(x).
\label{eq:smasses} 
\eea
These masses are fixed in terms of the overall scale parameter 
$\Lambda  \equiv \vevFS/\vevS$ and depend only very mildly on the mass ratios 
$x=\Lambda/\Mess{} $. $C_3$ equals 4/3 for squarks and 0 for sleptons, $C_2$ 
equals 3/4 for SU(2) doublets and 0 for singlets, and $Y = Q - T_3$.
In most of the parameter space that we analyze in the search of light
charged-Higgs and large $\tan \beta$ we find that $f(x),g(x) \simeq 1$.
 
The allowed values for the pseudoscalar Higgs boson mass $m_A$ in MGM 
theories with radiative symmetry breaking can be found from the tree-level 
formula at the electroweak scale, \eq{TLma}.
From Eqs.~(\ref{eq:gmasses}) and (\ref{eq:smasses}) we can relate
$\mbar{{H_u}}$ and $\mbar{{H_d}}$ with gaugino masses at the messenger scale:
\bea
\mbar{{H_u}}^2 = \mbar{{H_d}}^2 \simeq {3 \over 2~N} \Mbar{2} + {6 \over 5~N}
\Mbar{1}
\label{relmHM2}  
\eea
After running to the electroweak scale, in the large $\tan \beta$ region
these masses receive a large contribution proportional to $\Mbar{3}$ due to
the stop and sbottom masses, while the dependence on $\Mbar{2}$ and $\Mbar{1}$
through $\mbar{{H_u}}^2$ and $\mbar{{H_d}}^2$ is reduced because of large 
top and bottom Yukawa couplings. In fact $m_{H_u}^2(M_w)$ and
$m_{H_d}^2(M_w)$ are mainly determined by the values of $\Mbar{3}$. 
Therefore we can expect $m^2_A \simeq (C^g_u - C^g_d) M^{2}_{3} + \dots$, with 
$C^g_u$ and $C^g_d$ being both $O(1)$ in the large $\tan \beta$ regime.  
In MGM the minimum value of $\Mbar{3}$ is fixed by the lower bound on the 
lightest stau mass: 
\beq 
m_{\tilde \tau_1}^2  \simeq \mstau^2-m_{\tau}\mu\tan\beta  \gtrsim (100 \rm{GeV})^2\,.
\label{eq:staubound} 
\eeq  
where from Eq.~(\ref{eq:smasses}) we have, $\mstau^2\simeq{6\over5}\Mhat1^2+
s^2_W  m^2_Z$. From here, we end up  with $\Mhat1\gtrsim 300~\rm{GeV}$ and
thus $\Mhat3\gtrsim 1350~\rm{GeV}$.
Given that $m^2_A \simeq (C^g_u - C^g_d)\Mhat3^2$, with  both $C^g_u$ and
$C^g_d$ $O(1)$, a $m_A \sim 200~ \rm{GeV}$ would require a strong cancellation 
between the two Yukawa-dependent coefficients at a level of 
$C^g_u - C^g_d \sim 10^{-2}$. This can be compared with the 
situation  in the CMSSM where  we can take $M_{1/2}$ of order 200 GeV and we 
only need $C^g_u - C^g_d \sim 1$.

This estimate is confirmed by our numerical analysis,  performed by means of a
scanning over the MGM parameter space with SPheno \cite{Porod:2003um}. In
particular, we have not been able to find points with $m_A < 300 \rm{GeV}$. 
Given that in the present analysis we are interested in scenarios allowing a
light-heavy Higgs sector with $m_A \lesssim 200 \rm{GeV}$, the phenomenology 
of MGM models will be not analyzed.

In Anomaly Mediation the SUSY breaking is transmitted from the hidden sector
by the the superconformal anomaly \cite{Randall:1998uk,Giudice:1998xp}.
All the  soft-breaking parameters are determined in  a renormalization group
invariant way by a single parameter, the gravitino mass. The soft-breaking
parameters are given by:
\bea 
\label{eq:pureanom} 
 M_a ~=~  \frac{1}{g_a}~ \beta_a ~ m_{3/2}\, , \qquad\qquad 
m_i^2~=  ~\frac{1}{2}~  \dot \gamma_i ~ m_{3/2}^2\, , \qquad\qquad 
 A_i~ =~ \beta_{Y_i}~ m_{3/2} \, ,
\eea  
 where $\beta_a$ and $\beta_{Y_i}$ are the beta functions of gauge and Yukawa
couplings,   $\gamma_i$ the anomalous dimension of the corresponding matter
 superfield  and $m_{3/2}$ the gravitino mass. Unfortunately, pure
anomaly   mediation  is not acceptable because it leads to tachyonic
sleptons. The different approaches to solve this problem  make the analysis 
highly model dependent. A simple solution to this
problem   maintaining also the renormalization group invariance is to add a (or
several) Fayet-Iliopoulos D-term contribution(s) to the scalar masses 
 \cite{Kitano:2004zd,Hodgson:2005en}.  In this way the scalar masses in
 \eq{eq:pureanom} are replaced by $ 
 m_i^2~=~  \frac{1}{2} ~\dot \gamma_i ~ m_{3/2}^2 + m_0^2 ~Y_i$,
 where $m_0$  is the D-term contribution and $Y_i$ the charge corresponding to
 the new broken $U(1)$ symmetry. The spectrum depends then  on the charges of
 the  SM particles  under the new $U(1)$ groups.  
Although we do not make a full analysis, following Ref.  
 \cite{Hodgson:2005en}, we can see that the requirement of  $m_A > 90$ GeV 
sets one of the limits on the allowed region of the parameter space. 
Therefore in these particular models it is relatively easy to have  
 pseudoscalar masses below 200 GeV. However, a correct electroweak symmetry 
breaking is obtained only for values of $\tan \beta < 27$ making most of the 
phenomenology and specially indirect searches less interesting 
\cite{Hodgson:2005en}. In Ref. \cite{Baek:2002wm} a different solution to the
tachyonic problem is  analyzed with similar results. A complete 
analysis of more general anomaly mediation models is indeed interesting and
will be  discussed elsewhere.
\section{Generic signals of the light charged-Higgs scenario} 
\label{sec:signals}

As we have seen in the previous section, it is still possible to have a light
charged Higgs both in CMSSM and in NUHM models consistent with all the
phenomenological constraints. At this point we can ask what would be the
signatures of this scenario. We will discus the possible signals both at 
high-energy colliders (LHC, Tevatron) and at low-energy flavour changing  
experiments. 

\subsection{Direct searches at colliders}

The expected spectrum in the light charged-Higgs scenario is somewhat 
peculiar. In Table \ref{tab:inputA} we present the allowed range of input 
parameters in the CMSSM for points with a charged-Higgs below 200 GeV 
satisfying all direct constraints and indirect constraints and
a dark matter abundance in the range $0.08 < \Omega_\chi < 0.14$. 
As we can see, the main features of this region of parameter space are 
$M_{1/2} << m_0$ and $\tan \beta > 55$. 
As a consequence, we can expect relatively light gauginos 
and heavy sfermions. This is confirmed in  Table \ref{tab:spectrumA} where we 
show the obtained mass ranges with these input parameters. 
In this table we see that sfermions of the first two generations are roughly 
above 1 TeV. Only sfermions of the third generation can be relatively light
due to the effect of the large Yukawa couplings. 

\begin{table}
\begin{center}
\begin{tabular*}{0.7\textwidth}{@{\extracolsep{\fill}}|| c  l ||} \hline
  {\rm CMSSM parameter} & {\rm Allowed range}~~~~~~~\\ \hline
$m_0$  & 900 -- 1400 GeV \\ \hline
$M_{1/2}$  & 320 -- 440 GeV \\ \hline
$A_0$  & 350 -- 1700 GeV \\ \hline
$\tan \beta$  & 55 -- 60 \\ \hline
sign($\mu$)  & +1 \\ \hline
\end{tabular*}
\caption{\label{tab:inputA}
Input ranges in the CMSSM for points with $m_A < 200$ GeV and satisfying
direct, indirect and dark matter constraints.} 
\end{center}
\end{table}

\begin{table} 
\begin{center}
\begin{tabular}{|c| c||c| c|} \hline
\hline
  &~~~~~~~~~~~~~~{\rm mass (GeV)}~~~~~~~~~~~~~~  &    &~~~~~~~~~~~~~~{\rm mass
    (GeV)}~~~~~~~~~~~~~~    
 \\ \hline 
~~$\chi_1$~~  &  130 -- 180  & ~~$\chi_2$~~   & 250 -- 330   \\ \hline
$\chi_3      $  &  430 -- 540 & $\chi_4      $ &  450 -- 550 \\ \hline
$\chi^{\pm}_1      $  &  250 -- 330 & $\chi^{\pm}_2$  & 450 -- 550 \\ \hline
$\tilde g$ &  820 -- 1050 &  &   \\ \hline
$\tilde t_1 $  &  780 -- 1050 & $\tilde t_2 $& 890 -- 1170  \\  \hline
$\tilde b_1 $  &  850 -- 1150 & $\tilde b_2 $& 930 -- 1200  \\ \hline
$\tilde u_R $  &  1160 -- 1550  & $\tilde u_L $&  1180 -- 1560  \\ \hline
$\tilde d_R $  & 1150 -- 1550  & $\tilde d_L $& 1170 -- 1570 \\ \hline
$\tilde \tau_1$  & 320 -- 860  & $\tilde \tau_2$  & 720 -- 1160   \\ \hline
$ \tilde e_R$  & 900 -- 1360   &$ \tilde e_L$ & 920 -- 1380 \\ \hline
$\tilde \nu_1 $  & 700 -- 1160  & $\tilde \nu_3 $& 920 -- 1380    \\ \hline
$h      $  & 112.4 -- 115.6  & $H      $  &  165 -- 200 \\ \hline  
$A      $  & 165 -- 200   & $H^{\pm}$  & 180 -- 210  \\  \hline 
\end{tabular}
\caption{\label{tab:spectrumA} 
Mass ranges in the CMSSM for the input parameters in Table \ref{tab:inputA}.} 
\end{center}
\end{table} 

\begin{table}
\begin{center}
\begin{tabular*}{0.7\textwidth}{@{\extracolsep{\fill}}|| c  l ||} \hline
\hline
  {\rm CMSSM parameter} & {\rm Allowed range}~~~~~~~\\ \hline
$m_0$  & 760 -- 1280 GeV \\ \hline
$m_{H_{d 0}}$  & 660 -- 1380 GeV \\ \hline
$m_{H_{u 0}}$  & 820 -- 1520 GeV \\ \hline
$M_{1/2}$  & 180 -- 480 GeV \\ \hline
$A_0$  & 400 -- 2150 GeV \\ \hline
$\tan \beta$  & 39 -- 60 \\ \hline
sign($\mu$)  & +1 \\ \hline
\end{tabular*}
\caption{\label{tab:inputB}
Input ranges in the NUHM for points with $m_A < 200$ GeV and satisfying
direct, indirect and dark matter constraints.} 
\end{center}
\end{table}

\begin{table} 
\begin{center}
\begin{tabular}{|c| c||c| c|} \hline
\hline
  &~~~~~~~~~~~~~~{\rm mass (GeV)}~~~~~~~~~~~~~~  &    &~~~~~~~~~~~~~~{\rm mass
    (GeV)}~~~~~~~~~~~~~~    
 \\ \hline 
~~$\chi_1$~~  &  65 -- 195  & ~~$\chi_2$~~   & 120 -- 370   \\ \hline
$\chi_3      $  &  160 -- 640 & $\chi_4      $ &  240 -- 640 \\ \hline
$\chi^{\pm}_1      $  &  110 -- 370 & $\chi^{\pm}_2$  & 240 -- 650 \\ \hline
$\tilde g$ &  480 -- 1140 &  &   \\ \hline
$\tilde t_1 $  &  710 -- 970 & $\tilde t_2 $& 870 -- 1120  \\  \hline
$\tilde b_1 $  &  840 -- 1100 & $\tilde b_2 $& 900 -- 1210  \\ \hline
$\tilde u_R $  &  1080 -- 1520  & $\tilde u_L $&  1080 -- 1540  \\ \hline
$\tilde d_R $  & 1070 -- 1520  & $\tilde d_L $& 1080 -- 1540 \\ \hline
$\tilde \tau_1$  & 200 -- 1060  & $\tilde \tau_2$  & 620 -- 1200   \\ \hline
$ \tilde e_R$  & 780 -- 1300   &$ \tilde e_L$ & 800 -- 1310 \\ \hline
$\tilde \nu_1 $  & 610 -- 1190  & $\tilde \nu_3 $& 800 -- 1310    \\ \hline
$h      $  & 112.4 -- 115.6  & $H      $  &  128 -- 200 \\ \hline  
$A      $  & 128 -- 200   & $H^{\pm}$  & 148 -- 210  \\  \hline 
\end{tabular}
\caption{\label{tab:spectrumB} 
Mass ranges in the NUHM for the input parameters in Table \ref{tab:inputB}.} 
\end{center}
\end{table}

In the NUHM the allowed range of input parameters is shown in Table 
\ref{tab:inputB}. The main difference between the CMSSM and NUHM input 
parameters is the allowed range of  $\tan \beta$ compatible with a
light charged-Higgs below 200 GeV. In fact, values of 
$\tan \beta\gsim 40$ are allowed in the NUHM case. This implies that lower 
$M_{1/2}$ values with respect to the CMSSM case are now allowed. Lower
$M_{1/2}$ implies lighter squarks and gauginos that increase the SUSY
contribution to $BR(B\to s \gamma)$. However in the NUHM scenario, this can 
be compensated by selecting lower $\tan \beta$ values.
The particle mass ranges in NUHM models are shown in Table
\ref{tab:spectrumB}.   The comments made for the CMSSM apply also
here.  Notice that the CMSSM is a particular case of the NUHM, so all the 
allowed points in the CMSSM are also allowed in the NUHM.

\begin{figure}[t]
\begin{center}
\hspace{-0.3 cm}
\includegraphics[scale=0.40]{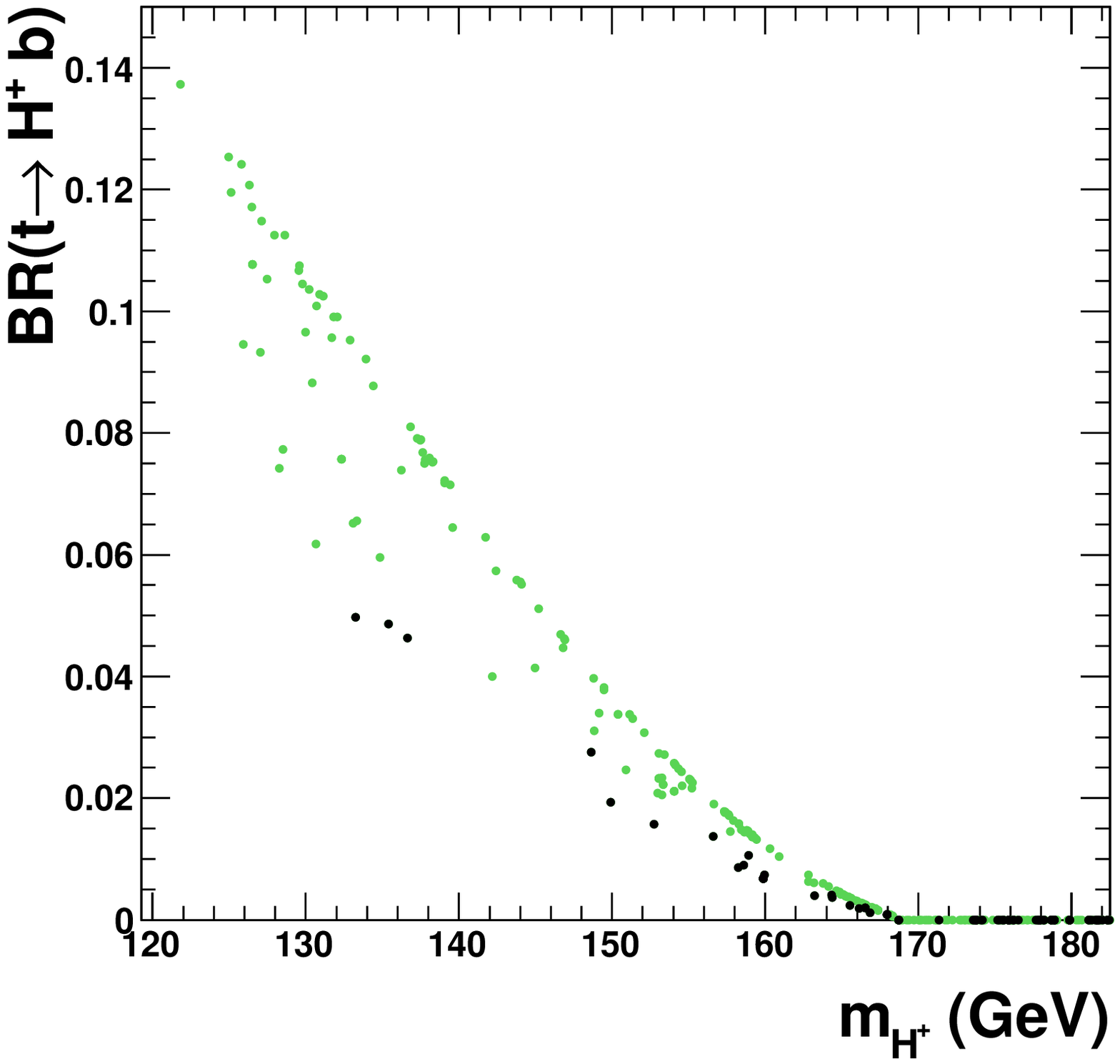}
\hspace{-0.3 cm}
\includegraphics[scale=0.40]{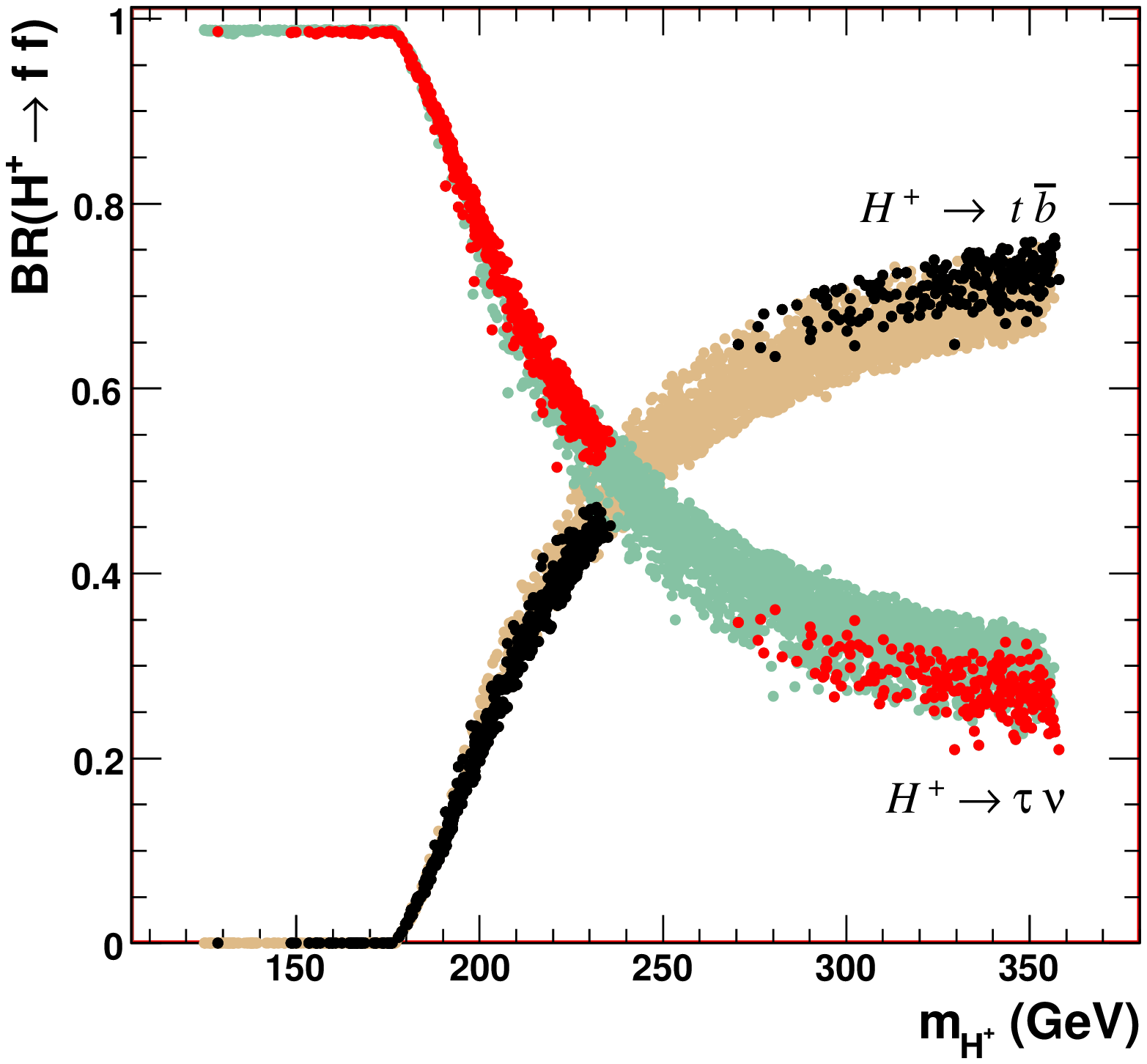}
\hspace{-0.3 cm}\\
\vspace{0.5 cm}
\caption{\label{fig:Higgs} 
Left: $BR(t\to H^+b)$ as a function of $m_{H^+}$.
Right: $BR(H^+\to t b)$ and $BR(H^+\to \tau \nu)$'s as function of $m_{H^+}$.
In both plots, clear (green or orange) points satisfy all the constraints on 
the SUSY spectrum from flavor and EWPO observables with the exception of 
$BR(B_u \to \tau \nu)$. Dark (red or black) points satisfy in addition the
constraints from  $BR(B_u \to \tau \nu)$.}
\end{center}
\end{figure}

Let us first discus the Higgs sector in this scenario. As can be seen in
Tables~\ref{tab:spectrumA} and \ref{tab:spectrumB}, both  the CMSSM and NUHM
predict the lightest Higgs boson in the range $m_h \simeq$ 112 -- 116 GeV.
Notice that our scenario corresponds, by construction, to the non-decoupling
regime of the MSSM. Hence the LEP bound on the SM Higgs mass of 114.4 
GeV does not apply in our case. In general values as low as 90 GeV for the
lightest Higgs mass are allowed in this regime \cite{Schael:2006cr}.
The lower value for the lightest Higgs mass we obtain is due to the heavy
squark masses and large $\tan \beta$ values. As we have discussed, we select
the points in the parameter space in order to get charged Higgs masses $\lsim 200$
GeV. This opens the possibility of interesting experimental signatures at
colliders. In particular, one of the more interesting possibilities
would be to look for the top decay via $t\to H^{+}b$, which is  allowed for 
$m_t >m_{H^+}+m_b$. Otherwise, when this decay is not allowed,
charged-Higgs decays are also interesting. The charged
Higgs boson decays mainly to \taunus ~ or to $t b$ depending on $m_{H^+} - m_t -m_b$.
In the left-hand side plot of Fig.~\ref{fig:Higgs} we show the 
$BR(t\to H^+b)$ as a function of the charged Higgs mass $m_{H^+}$ imposing all  
the constraints on the SUSY spectrum from flavor and EWPO observables. 
We can see that branching ratios at the few per cent level are possible.  
Given that the LHC
will be a top factory producing $10^7$ top pairs already with 10 $fb^{-1}$, 
the $t\to H^+b$ process clearly represents a very clean signature
of our scenario if $m_t >m_{H^+}+m_b$, which is only possible in the NUHM 
model. Remember that in the CMSSM $\tan \beta > 53$ for $m_{H^+}<200$ GeV and 
then the constraints from $\Btaun$ forbid completely this possibility.  
Charged Higgs decays are also interesting in general. 
In the right-hand side plot of Fig.~\ref{fig:Higgs} we report the 
$BR(H^+\to t b)$ and   
$BR(H^+\to \tau \nu)$'s as function of $m_{H^+}$. We note that 
$\Gamma(H^+\to\rm{All})\simeq \Gamma(H^+\to tb)+\Gamma(H^+\to \tau \nu)$) 
and that $BR(H^+\to tb)$ increase while increasing $m_{H^+}$
as it is understandable by kinematical considerations.
On the other hand, the $H^+\to \tau \nu$ decay mode starts being
the dominant one when $m_{H^+}\leq 220 \rm{GeV}$.

As discussed above, our scenario predicts relatively light gaugino masses. In
particular the gluino mass turns out to be usually lighter than the squarks.
This has important phenomenological consequences at hadronic colliders.
In fact from Tables~\ref{tab:spectrumA} and \ref{tab:spectrumB} we can see
that the gluino is always lighter than squarks of the first two generations
thus the decay $\tilde g \to q_{1,2} \tilde q_{1,2}$ is never allowed. 
In general this is not always true for squarks of the third generation as can
be seen in the tables. However, in our numerical analysis we find that the
decay  $\tilde g \to t \tilde t_{1}$ is never kinematically allowed. On the
contrary,  $\tilde g \to b \tilde b_{1}$ is possible for $\lsim 10\%$ of the
allowed points. Notice that this fact is only due to the small bottom mass compared
to the top mass. If none of these decays is kinematically allowed, gluino has
either three  body decays into two quarks
and either a neutralino or a chargino or the loop induced two-body decay into
$g \tilde \chi^0_i$.  In the scenarios discussed above $\tan\beta$ is
relatively large implying that final states containing quarks of the 3rd
generations are strongly  preferred \cite{Bartl:1994bu}. The branching ratios
of the final states $b \bar{b} \tilde \chi^0_i$ are enhanced compared to $t
\bar{t} \tilde \chi^0_i$ due  to obvious kinematical reasons.Therefore,
independent of kinematics we expect in gluino decays an enhancement of final
states containing b-quarks.

\subsection{Indirect FCNC searches}

The main features of our scenario are large $\tan \beta$ and light Higgs
masses. Therefore we can expect sizeable SUSY contributions to $\tan
\beta$-enhanced decays, specially in $B_s\to\mu^+\mu^-$ and
$B\to\tau^+\nu^-$. Even though we find very large contributions to the above
observables in most of the points of our numerical analysis, we must stress 
that it is not guaranteed that an experimental/theoretical improvement in
these decays would find a departure from the SM expectations. In fact, in the
$B_s\to\mu^+\mu^-$ case we can always find regions where the smallness of
$A_t$ and/or gluino-chargino cancellations can reduce SUSY contributions to
the level of the SM. With respect to the $B\to\tau^+\nu^-$ decay we stress
that the light-Higgs scenario with large $\tan \beta$ values can accommodate
the present experimental determination only when the SUSY contribution is
roughly twice the SM one (note that the charged-Higgs contribution has always
opposite sign compared to the SM one). Therefore, even if tuned, it is always
possible to find a ($\tan \beta$, $m_{H^+}$) combination reproducing the SM
prediction for this branching ratio. However we must emphasize that both
decays are probably the most promising indirect channels to look for the light
charged-Higgs scenario.

In addition to these hadronic observables, lepton flavour violating (LFV)
transitions, as $\ell_i \rightarrow \ell_j\gamma$, are also very sensitive
probes of the large $\tan \beta$ scenario.  Unfortunately these decays require
an additional source of LFV. However, LFV couplings naturally appear in the
MSSM once we extend it to accommodate the non-vanishing neutrino masses and
mixing angles by means of a supersymmetric seesaw mechanism~\cite{fbam,hisano}.
In this case, LFV entries in the slepton mass matrix $(m^2_{\tilde{L}})_{ij}$
are radiatively induced \cite{fbam}:
\be
 \delta_{LL}^{ij} ~=~
\frac{ \left( M^2_{\tilde \ell} \right)_{L_i L_j}}
 {\sqrt{\left( M^2_{\tilde \ell} \right)_{L_i L_i}
\left( M^2_{\tilde \ell} \right)_{L_j L_j}}} \approx
 -\frac{(3+A_0^2)}{8\pi^2}\log\left(\frac{M_X}{M_R}\right)
(Y^\dagger_\nu Y_\nu)_{ij}~,
 \label{eq:fbam}
\ee
where $Y_\nu$ are the neutrino Yukawa couplings (the potentially large sources
of LFV) and $M_X$ and $M_R$ are the GUT and the heavy right handed neutrino
masses, respectively.
In our analysis, we consider a rather conservative situation where the mixing
angles in the neutrino Yukawa matrix are small, CKM-like \cite{Masierorew},
and the largest neutrino Yukawa eigenvalue is $O(1)$ similarly to the top Yukawa.

\begin{figure}[t]
\centering
\includegraphics[scale=0.40]{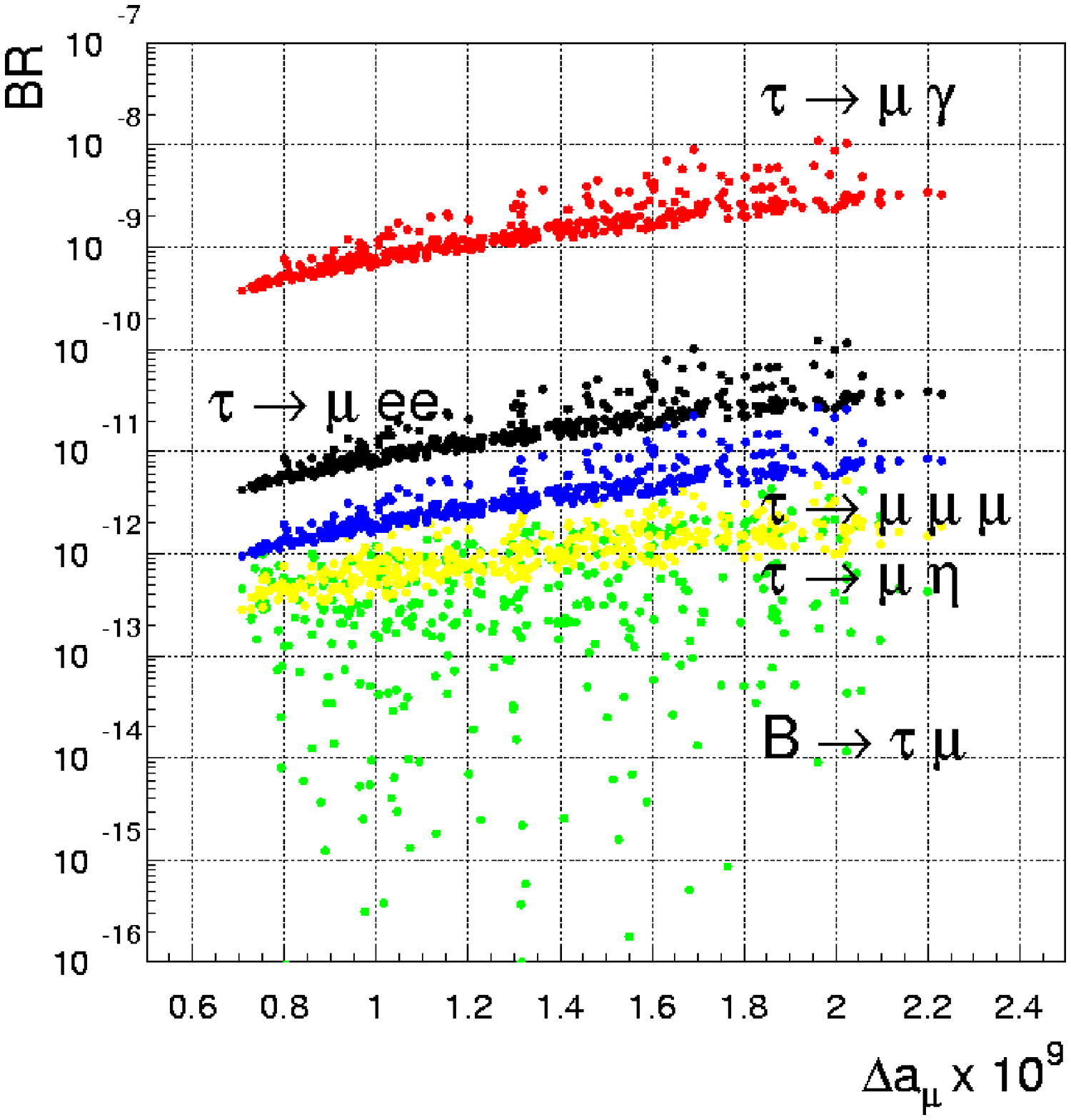}
\includegraphics[scale=0.40]{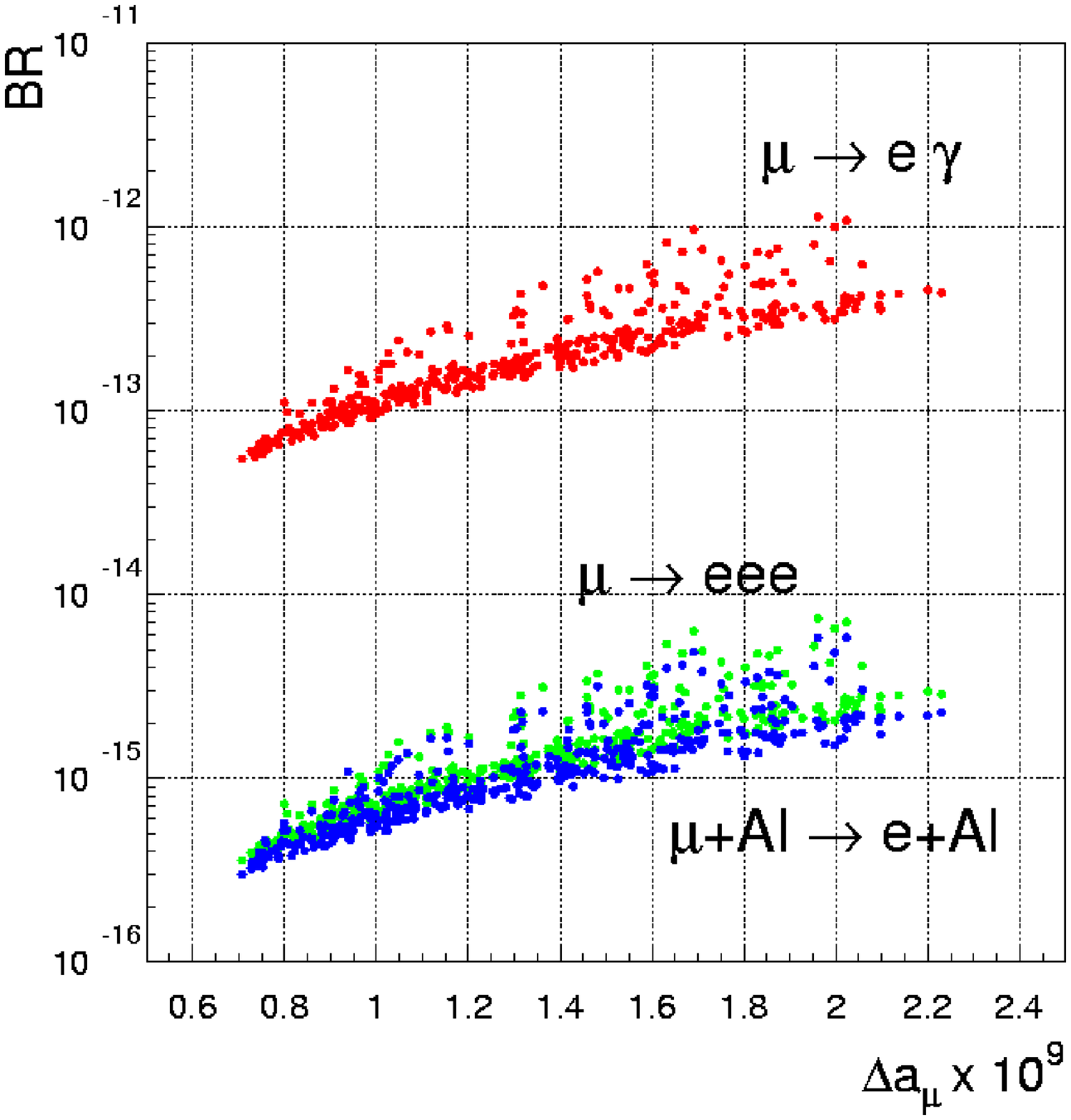}
\caption{\label{fig2}
Left (Right): Expectations for $\tau\to\mu$ ($\mu\to e$) transitions
vs.~$\Delta a_\mu=(g_\mu-g^{\rm SM}_\mu)/2$ (remind that 
$\Delta a_{\mu} = a_{\mu}^{\rm exp} - a_{\mu}^{\rm SM}\approx(3 \pm 1)\times 10^{-9}$)
assuming a CKM-like scenario. The plots have been obtained employing the ranges for
the input parameters listed in Table~\ref{tab:inputA}.
}
\end{figure}

In Fig.~\ref{fig2} on the left-(right-)hand side, we report the predictions 
of the CKM-like scenario for $\mu\rightarrow e$ ($\tau\rightarrow\mu$) 
transitions as a function of $\Delta a_{\mu}$ employing the ranges for the
input parameters listed in Table~\ref{tab:inputA}. We
set $y_{\nu_3}=1$, $M_X=2\times 10^{16}$ GeV and $M_R= 10^{15}$ GeV, as it would be
obtained via the see-saw formula with a hierarchical light neutrino spectrum
with $m_{\nu_1}\simeq 10^{-3}$ eV. Notice that in this figure we present the
predictions for LFV processes in the CMSSM scenario; in fact, within
the region of parameter space of our interest, the CMSSM and NUHM models
(with RH neutrinos) have very similar predictions. Given that both $\ell_{i}\rightarrow\ell_{j}\gamma$ and $\Delta a_\mu=(g_\mu-g^{\rm SM}_\mu)/2$
are generated by dipole operators, it is natural to expect that their amplitudes
are closely connected \cite{Hisa1,IP2}. In particular, assuming a CMSSM spectrum,
it is found that
\bea
\BR(\ell_i\rightarrow \ell_j \gamma) ~\approx~
\left[\frac{\Delta a_{\mu}}{ 30 \times 10^{-10}}\right]^{2}
\times \left\{\ba{ll} 10^{-12} \, 
\bigg| \frac{\delta_{LL}^{12}}{4\times 10^{-5}} \bigg|^2
\qquad  & [\mu\to e]~,\\ 10^{-8} \, 
\bigg| \frac{\delta_{LL}^{23}}{6\times 10^{-3}} \bigg|^2
& [\tau\to\mu]~.
\ea
\label{lfvgm2}
\right.
\eea
where $\delta_{LL}^{ij}$ has been evaluated by means of Eq.~(\ref{eq:fbam})
for $A_0 = 1$.
As we can see, the correlation is not exactly a line as one would expect from
Eq.~(\ref{lfvgm2}), since i) the loop functions for the two processes are not
identical, ii) while $BR(\ell_i \rightarrow \ell_j\gamma)$ strongly depends
on $A_0$ through $\delta_{LL}^{ij}$ (see Eq.~(\ref{eq:fbam})), $\Delta a_\mu$
is almost insensitive to $A_0$.
From Fig.~\ref{fig2} we see that, although model dependent, both
$\mu\rightarrow e\gamma$ and $\tau\rightarrow \mu\gamma$ branching ratios
could naturally reach the experimentally projected sensitivities in MEG
and SuperB factories. This is specially true in the interesting region of
the SUSY parameter space where also the $(g-2)_{\mu}$ anomaly, i.e.
$\Delta a_{\mu} = a_{\mu}^{\rm exp} - a_{\mu}^{\rm SM}
\approx(3 \pm 1)\times 10^{-9}$, can find a natural explanation.

Furthermore, we note that, although our scenario has a very light heavy Higgs
sector $m_A \lesssim 200~\rm{GeV}$ and large $\tan\beta$ values,
$\tan\beta\gtrsim 40$, Higgs mediated LFV transitions 
\cite{bkl,sherH,ellisH,okadaH,rossiH,paradisiH} are not particularly enhanced,
as could be expected.
The reason for this is that Higgs mediated lepton flavour violating couplings
are quite suppressed because of the large mass splitting between the gaugino
and sfermion masses. In particular, as we can see from Fig.~\ref{fig2}, both
$BR(\tau\to\mu\eta)$ and $BR(B\to\tau\mu)$ (that are purely Higgs mediated
processes \cite{bkl}), never exceed the level of few $10^{-12}$.
Higgs mediated contributions to $\tau\to\mu\mu\mu$ are completely subdominant
compared to the dipole ($\tau\to\mu\gamma^*$) effects.

As it concerns the $\mu\to e$ transitions, scalar current effects contribute quite
sizably only to $BR(\mu+ Al\to e + Al)$. However, for the parameter space relevant
in our analysis, both $BR(\mu+ Al\to e+Al)$ and $BR(\mu\to eee)$ lie below the
$10^{-14}$ level, well far from their current experimental resolutions.

Finally, the predictions for $\tau\to e$ transitions are simply obtained from
those for the $\tau\to\mu$ transitions by $BR(\tau\to e X)=|V_{td}/V_{ts}|^2
BR(\tau\to eX)$ (with $X=\gamma,\eta,\mu\mu,ee$) and $BR(B\to\tau e)=|V_{td}/V_{ts}|^2
BR(B\to\tau\mu)$.

\section{Conclusions and outlook}
\label{sec:conclusions}
In this paper we have investigated whether the possibility of having a 
charged-Higgs boson with a mass below 200 GeV is still open.
We have answered this question both in the context of a 2HDM and in the 
framework of the MSSM. In the 2HDM, the charged-Higgs mass is constrained 
to be above 295 GeV by BR$(b\to s \gamma)$ \cite{Misiak:2006ab},
although we have found that a pseudoscalar mass in the range 150--200 GeV is
still allowed. In the context of the
MSSM  we have seen that a light charged-Higgs below 200 GeV is still possible
both in the Constrained MSSM (CMSSM) and in a MSSM with non-universal Higgs masses.
Light Higgs masses in these supersymmetric scenarios require always very large
values of $\tan \beta$. These models, in the light $m_{H^+}$--large $\tan \beta$
region have to face strong restrictions from the $B \to \tau \nu$ decay, that turns
out to be the strongest constraint of our scenario. In particular, in the CMSSM
$\tan\beta$ is always larger than 50 when we want $m_{H^+}<200$ GeV and then the
$B\to\tau\nu$ decay sets a strict lower limit of 180 GeV for the charged-Higgs mass.
This lower limit from $B\to\tau\nu$ is relaxed in NUHM models where we can
obtain  light charged Higgses with smaller values of $\tan\beta$.
Moreover, we have analyzed the generic predictions of our light charged-Higgs
scenario for hadronic colliders and indirect searches.
Finally, we have addressed the question whether the above scenario can be
tested through LFV processes. To this purpose, we have considered a rather
conservative ansatz for the source of LFV, the so-called CKM-like
\cite{Masierorew} case, and we have evaluated the predictions for the most
relevant low-energy LFV processes.
Interestingly enough, both $\mu\rightarrow e\gamma$ and 
$\tau\rightarrow \mu\gamma$ branching ratios naturally reach the
experimentally projected sensitivities in MEG and SuperB factories, specially
in the region where the $(g-2)_{\mu}$ anomaly can find a natural explanation.

\section*{Acknowledgments}

We thank M.Antonelli, U.Haisch, G.Isidori, J.Hisano, M. Carena and C. Wagner 
for useful discussions.
G.B., P.P. and O.V. acknowledge partial support from the Spanish 
MCYT FPA2005-01678, Generalitat Valenciana under contract GV05/207 and the EU
MRTN-CT-2006-035482 (FLAVIAnet). 
W.P.~is partially supported
by the German Ministry of Education and Research (BMBF) under
contract 05HT6WWA. E.L.~is supported by the Department of Energy under Grant 
DE-AC02-76CH030000. E.L and P.P thank the Aspen Center for Physics, where part
of this work was done.
 Fermilab is operated by Fermi Research Alliance, LLC under Contract 
No. DE-AC02-07CH11359 with the United States Department of Energy.

\end{document}